%% file: NA62_ee_accepted.tex
\documentclass[
superscriptaddress,
preprint numbers,
amssymb,
nofootinbib,
twocolumn,
]{revtex4-2}

\usepackage{bm}
\usepackage{esvect}
\usepackage{verbatim}
\usepackage{multirow}

\usepackage[left]{lineno}
\usepackage{blindtext}
\pdfoutput=1
\usepackage[T1]{fontenc}
\usepackage{graphicx}
\usepackage{epsfig}
\usepackage{amsmath}
\usepackage{amsfonts}
\usepackage{amssymb}
\usepackage{braket}
\usepackage{cancel}
\usepackage{pstricks}
\usepackage{color}
\usepackage{xcolor}
\usepackage{feynmf}
\usepackage[normalem]{ulem}
\usepackage{epstopdf}
\usepackage{soul}
\usepackage{cancel}

\usepackage[footnote]{witharrows}
\usepackage{braket}
\usepackage{cancel}
\usepackage{setspace}
\usepackage{pstricks}
\usepackage{xcolor}
\usepackage{float}
\usepackage{mathtools}
\usepackage{multirow}
\usepackage{booktabs}
\usepackage{enumitem}

\newcommand{\darkphoton}{A^{\prime}}

\begin{document}

\preprint{}
\title{ 
{\normalfont{\Large EUROPEAN ORGANIZATION FOR NUCLEAR RESEARCH} \vspace{3mm} 
\begin{flushright}
\end{flushright}
}
\vspace{40mm}
Search for Leptonic Decays of Dark Photons at NA62
}

\author{The NA62 collaboration}
\affiliation{Full author list given at the end of the Letter.}


\date{\today}

\begin{abstract}
The NA62 experiment at CERN, configured in beam-dump mode, has searched for dark photon decays in flight to electron-positron pairs using a sample of $1.4\times 10^{17}$ protons on dump collected in 2021.
No evidence for a dark photon signal is observed. The combined result for dark photon searches in lepton-antilepton final states is presented and a region of the parameter space is excluded at 90\% CL, improving on
previous experimental limits for dark photon mass values between 50 and 600~MeV$/c^2$ and coupling values in the range $10^{-6}$ to $4\times10^{-5}$. 
An interpretation of the $e^+ e^-$ search result in terms of the emission and decay of an axion-like particle is also presented.
\end{abstract}

\maketitle
\flushbottom

\section{Introduction}
\label{eq:intro}
The prevalence of dark matter over ordinary matter, 
one of the unsolved puzzles of the universe, has inspired various extensions of the Standard Model (SM). Some of these predict the existence of an additional $U(1)$ gauge-symmetry sector with a vector mediator field $\darkphoton$ known as a ``dark photon''.

Minimalistic dark photon models~\cite{Okun, Holdom} introduce the $A^\prime$ field with mass $M_{A^\prime}$, which interacts with the gauge field $B$ associated with the SM $U(1)$ symmetry through kinetic mixing, with its strength characterised by the coupling constant $\varepsilon$. 
The dark photon may also interact with additional fields in the dark sector. Under the assumption that $M_{A^{\prime}}$ is lower than twice the mass of the lightest state in the dark sector, the dark photon decays to SM particles only. Cosmological constraints on the thermal relic density of dark matter favour a dark photon mass range from 1 to 1000~MeV/$c^2$, together with $\varepsilon$ within the range $10^{-6}$ to $10^{-3}$~\cite{Boehm_2004,Pospelov_2008,Feng_2008}. In this range of parameters, the decay length of dark photons with momenta exceeding 10~GeV/$c$ vary from tens of centimetres to hundreds of metres. For masses below 700~MeV/$c^2$, the primary contribution to the dark photon decay width arises from di-lepton final states~\cite{Batell_PRD2009}. Dark photon searches in beam-dump experiments exhibit superior sensitivity within the above parameter space region compared to searches at colliders or in meson decays. An extensive survey of the experimental methods is presented in~\cite{Reece_JHEP_2009}. 

Proton interactions in the dump can produce dark photons by two mechanisms, bremsstrahlung and decays of secondary neutral mesons. The former case is interpreted as a scattering process in the Fermi-Weizs\"acker-Williams approximation~\mbox{\cite{Blumlein_PLB2014}} where a virtual photon is exchanged between the primary proton and a nucleus, resulting in a dark photon and a scattered proton in the final state. In the latter process, a dark photon is emitted alongside a photon or a neutral meson~\mbox{\cite{Blumlein_PLB2011}}. Decays of $\pi^0, \eta,\eta^{\prime},\rho, \omega, \phi$ mesons are relevant for this analysis.

The search for a dark photon decaying into $e^+ e^-$ is described here.
The result of this search in combination with a previous result~\cite{Dobrich:2023dkm} for the $\darkphoton\to\mu^+ \mu^-$ decay is presented.

Axion-like particles (ALPs) are hypothetical
pseudoscalar particles arising in many extensions of the SM. A scenario of emission of an ALP coupled to the SM fermionic fields is also considered.
In proton-nucleus collisions, an ALP $a$ can be produced in the decays of charged and neutral $B$ mesons as \mbox{$p N \to B X, ~\mbox{followed by }B\to K^{(*)}a$}~\cite{Dobrich_PLB2019}, where $K^*$ is the $K^*(892)$ resonance.
In this work, a general scenario where the coupling of ALPs to SM fermions is not uniform (meaning the coupling to leptons could be different from the coupling to quarks) is addressed.

\section{BEAMLINE AND DETECTOR}
\label{sec:setup}

\begin{figure*}[ht]
    \centering
    \includegraphics[width=0.8\linewidth]{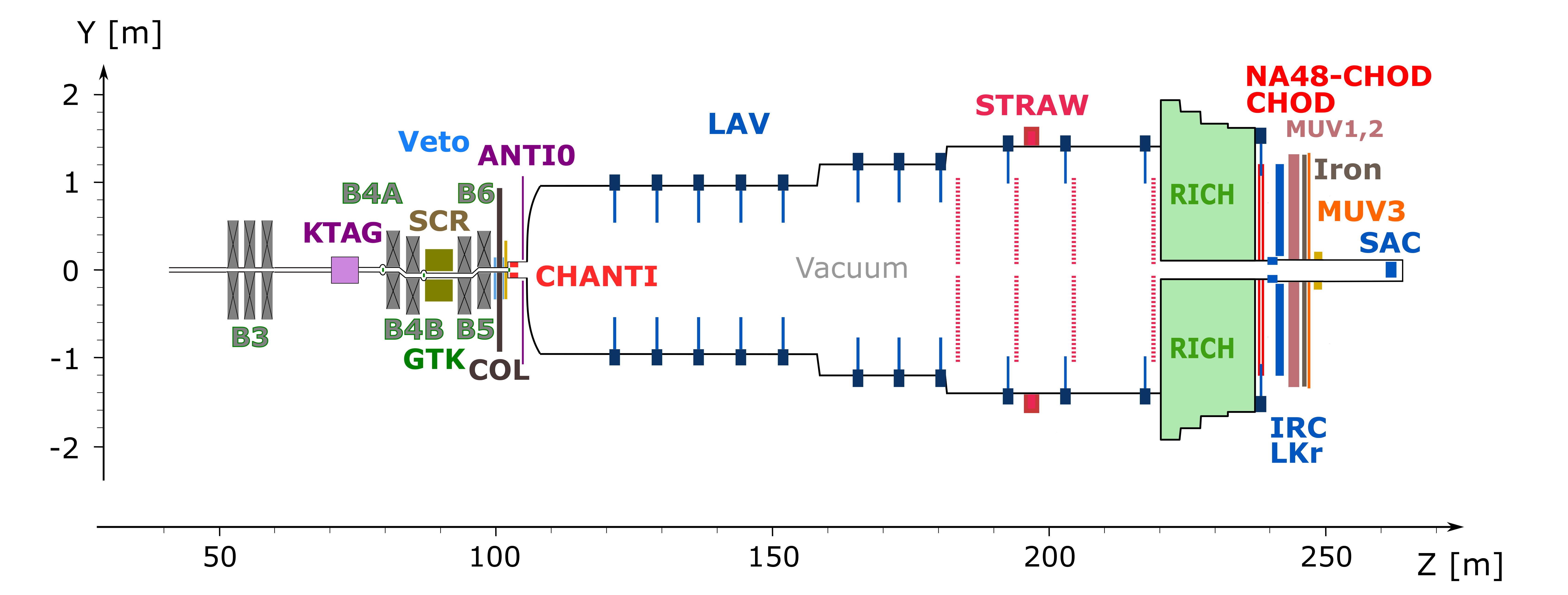}
    \caption{Schematic side view of the NA62 setup in 2021. Information from KTAG, GTK, CHANTI, MUV1,2, IRC, and SAC is not used in this analysis. Not all beam elements are shown.}
    \label{fig:na62_layout}
\end{figure*}

Figure~\ref{fig:na62_layout} illustrates the NA62 beamline and detector layout. A comprehensive description of these components can be found in~\cite{na62det}. 
In standard operation, kaons are produced by 400 GeV/$c$ protons extracted from the CERN SPS impinging on a beryllium target. In dump-mode operation, the beryllium target is removed, and the protons interact in a 3.2~m long absorber (TAX) equivalent to 19.6 nuclear interaction lengths. 
The origin of the coordinate system is in the centre of the target. The Z axis points in the proton beam direction, the Y axis points upwards, and the X,Y,Z axes form a right-handed system. The mean position of the primary protons at the TAX entrance is $(0, -22\mathrm{~mm}, 23\mathrm{~m})$.

The momenta and directions of charged particles within the fiducial volume (FV) are measured by a magnetic spectrometer (STRAW). A quasi-homogeneous liquid krypton electromagnetic calorimeter (LKr) and a muon detector (MUV3) are used for particle identification. Twelve ring-shaped lead-glass detectors (LAV1–12) record activity originating from secondary interactions. 
Two scintillator hodoscopes, NA48-CHOD and CHOD, provide trigger signals and time measurements for charged particles with 200~ps and 800~ps resolution, respectively. The  ANTI0 scintillator hodoscope~\cite{Danielsson_2020} is used to detect charged particles produced upstream of the FV. Further details of the beam-dump mode operation are given in~\cite{Dobrich:2023dkm}.

\section{ANALYSIS STRATEGY AND EVENT SELECTION}
\label{sec:analysis}
The search is based on the data sample collected in a 10-day period in 2021, corresponding to $1.4 \times 10^{17}$ protons on TAX (POT). Three trigger lines were implemented: Q1 required at least one signal in the CHOD, downscaled by a factor of 20; H2 required in-time signals in two CHOD tiles; and a control trigger required an  LKr energy deposit above $1$~GeV.  

Two-track final states triggered by the H2 condition are considered. Each track reconstructed by the STRAW must satisfy the following criteria: momentum $p> 10$~GeV/$c$; extrapolated positions at the front planes of NA48-CHOD, CHOD, LKr, MUV3 within the geometrical acceptance of each detector; extrapolated positions at the first STRAW chamber and LKr front planes isolated from those of other tracks. Each track must be associated with a CHOD signal compatible in space and time. 
The track time is defined using the time of the associated NA48-CHOD signal if present, otherwise using the time of the associated CHOD signal. 
Track times must be within 5~ns of the trigger time. Tracks spatially compatible and in-time with an ANTI0 signal or in-time with a LAV signal are rejected.

Any MUV3 signal within a momentum-dependent search radius around the extrapolated track position and within 5~ns of the track time is associated with the STRAW track. 
An LKr energy deposit $E >1$~GeV is associated with the track if it is in-time and spatially compatible, accounting for possible bremsstrahlung-induced energy deposits. 
Tracks with an associated MUV3 signal and $E/p < 0.2$ are identified as muons. Tracks without associated MUV3 signals, with $(E/p)_{\rm{min}} < E/p < 1.05$ are identified as electrons, where $(E/p)_{\rm{min}} = 0.95$ for $p < 150$~GeV/$c$ and decreases with momentum otherwise.

\begin{figure}[htb]
    \centering
    \includegraphics[width=\columnwidth]{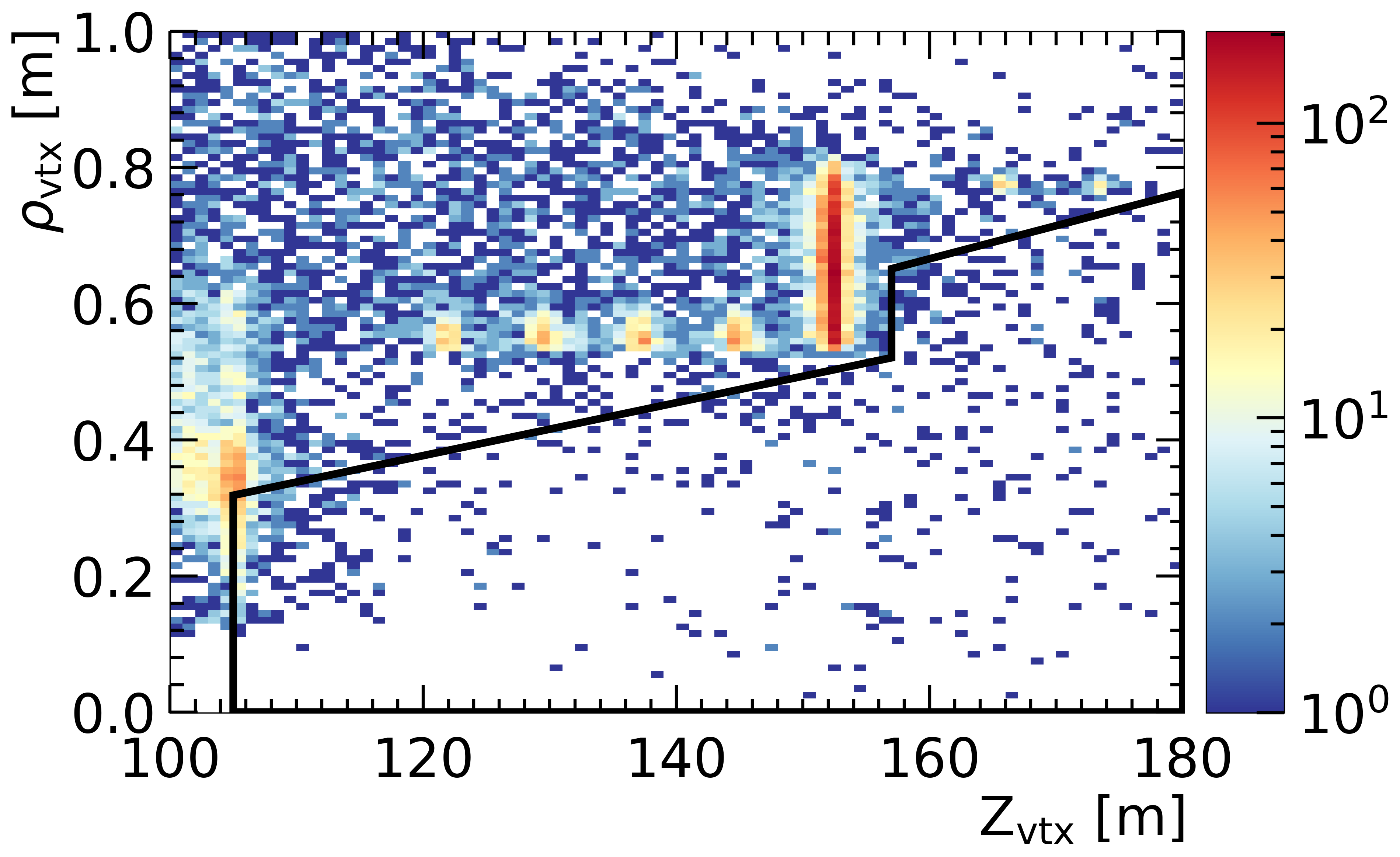}
    \caption{Distribution of two-track vertex positions in the plane ($Z_{\mathrm{vtx}}$, $\rho_{\mathrm{vtx}}$) for data events, without particle identification requirements. The black contour defines the restricted FV.}
    \label{fig:FV_def}
\end{figure}

Two time-coincident tracks consistent with originating from a common point form a vertex. The presence of exactly one two-track vertex is required, regardless of the total number of tracks in the event. The vertex time is evaluated as the mean time of the two tracks. The vertex position is obtained by the backwards extrapolation of the tracks, accounting for the residual magnetic field in the FV. The data distribution of the vertex longitudinal coordinate ($Z_{\mathrm{vtx}}$) and radial position in the transverse plane ($\rho_{\mathrm{vtx}}$) is shown in Figure~\ref{fig:FV_def}, without the particle identification (PID) criteria applied. This distribution is dominated by secondary interactions in LAV1--5 and in the front vacuum-tank window. Most reconstructed vertices originate from secondary interactions in LAV5 ($Z\simeq 152$~m). LAV6--12 have larger inner radii (Figure~\ref{fig:na62_layout}) and do not block the resulting particles. It is required that the vertex is reconstructed in the restricted FV, defined as shown in Figure~\ref{fig:FV_def}, to reject these interactions.  


The position of the $\darkphoton$ production point is evaluated as the point of closest approach between the $\darkphoton$ line of flight, defined by the two-track vertex position and total momentum direction, and the beam line, parallel to the Z axis and defined by the average impact point of the primary protons in the TAX. 
The signal region (SR) is defined as an ellipse in the plane of the Z coordinate $(\mathrm{Z}_{\mathrm{TAX}})$ and the distance between the two lines $(\mathrm{CDA}_{\mathrm{TAX}})$:
\begin{equation}\label{eq:ImprovedSR}
    \mathrm{SR:}\,\, 
    \left(
    \frac{\mathrm{Z}_\mathrm{TAX}[\mathrm{m}] - 23}{12}
    \right)^2 + 
    \left(
    \frac{\mathrm{CDA}_\mathrm{TAX}[\mathrm{m}]}{0.03}
    \right)^2  < 1.
\end{equation}
This condition reduces the signal acceptance by 1.7\% as shown by simulation.
The control region (CR) used to validate the background estimate is the area outside SR that satisfies:
\begin{equation}\label{eq:CR}
    \mathrm{CR:}\,\,-4<\mathrm{Z}_\mathrm{TAX} <50~\mathrm{m}\,\text{ and }\,\mathrm{CDA}_{\text{TAX}}<0.15~\mathrm{m}. 
\end{equation}

\begin{figure}[htb]
    \centering
    \includegraphics[width=\columnwidth]{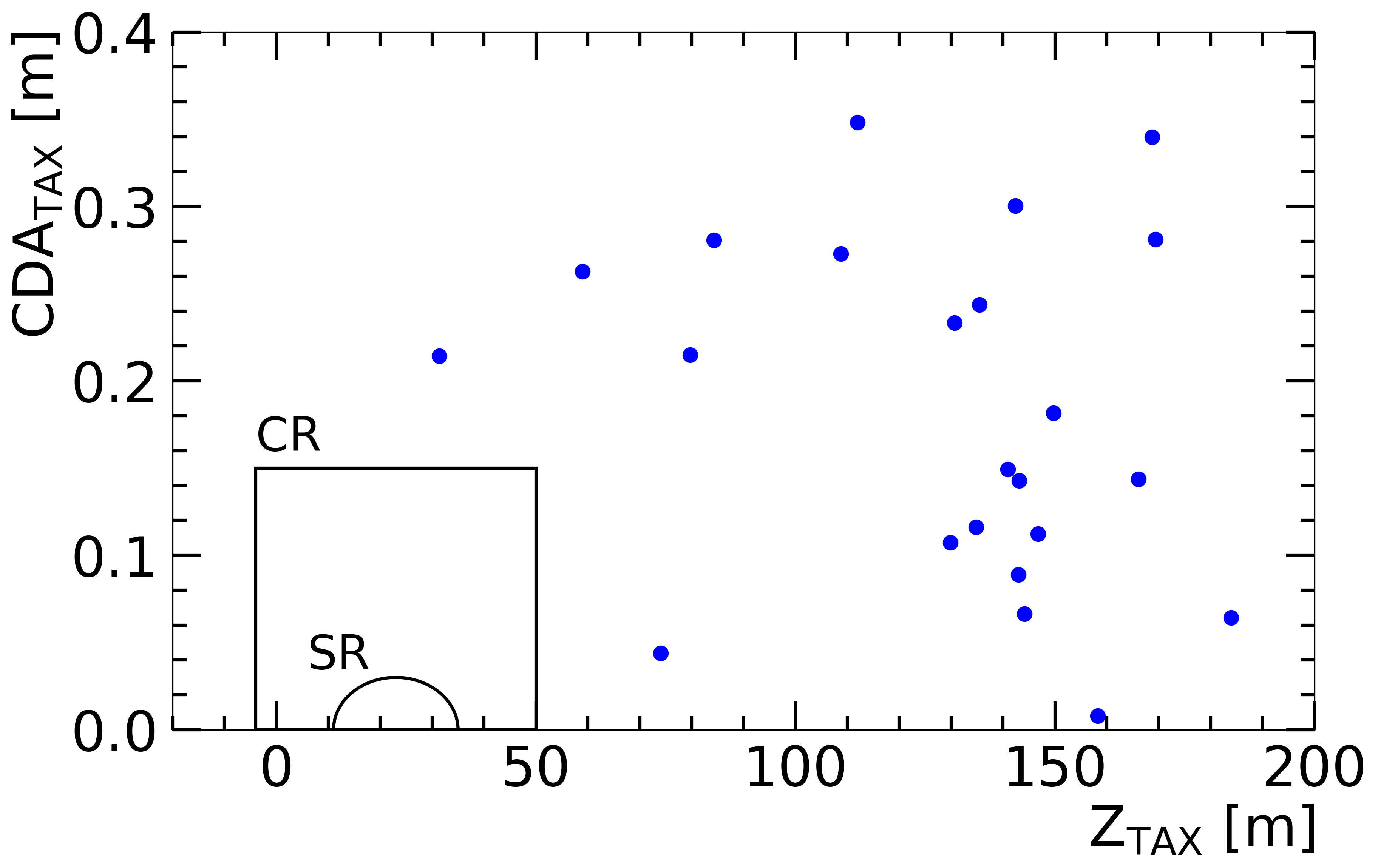}
    \caption{Data distribution in the plane  $(\mathrm{Z}_{\mathrm{TAX}},\mathrm{CDA}_{\mathrm{TAX}})$ for $e^+e^-$ vertices without applying the LAV and ANTI0 veto conditions. CR and SR are masked.}
    \label{fig:cda_vs_z_tax}
\end{figure}

Both SR and CR are kept masked until validation of the background estimate. The data distribution of $e^+e^-$ vertices in the plane $\left(\mathrm{Z}_{\mathrm{TAX}},\mathrm{CDA}_{\mathrm{TAX}}\right)$, after applying the full selection except for the LAV and ANTI0 veto conditions, is shown in Figure~\ref{fig:cda_vs_z_tax}. The full selection removes all events outside SR and CR. 

\section{BACKGROUNDS}
\label{sec:bkg}
Mesons produced by proton interactions in the TAX generate a flux of ``halo'' muons. 
The dominant background involves vertices in which both particles are created by the same halo muon interacting with the material along the beamline (prompt background). A control data sample is constructed from muons satisfying the Q1 trigger and not the H2 trigger. These muons are extrapolated backwards using PUMAS~\cite{PUMAS} to the upstream plane of the B5 magnet (Figure~\ref{fig:na62_layout}) and are used as input to a \verb|GEANT4|-based Monte Carlo (MC) simulation~\cite{geant4}. The resulting events constitute the prompt background sample, which is subjected to the signal selection. The size of this control sample is equivalent to that of the data sample. The expected number of $e^+e^-$ vertices reconstructed in the restricted FV derived from this sample has a relative systematic uncertainty of $50\%$ arising from the limited accuracy of the backward extrapolation and forward propagation.

Another possible source of background is the random pairing of $e^+$ and $e^-$ tracks originating from different primary proton interactions. The combinatorial background component is evaluated using a data-driven approach, with events triggered by the Q1 condition. This approach considers all possible sources of single electrons, including decays and interactions of secondary mesons occurring in the FV, or close to its boundary. Single tracks are paired within a 10~ns time window, building pseudo-events. For each event, the vertex is reconstructed as in the signal selection, and the event is assigned a weight that accounts for the time window and the downscaling factor of the Q1 trigger. This background is found to be an order of magnitude smaller than the prompt background, and is therefore neglected.

Backgrounds from neutrino interactions and $K_L$ decays are negligible with respect to the prompt background. Tracks with PID other than $e^+$ or $e^-$ have been reconstructed in the data sample. These extrapolate backwards either to one of the LAV stations (excluded from the FV definition) or to the upstream region and are vetoed by the ANTI0. Finally, the number of $\mu\pi$, $\mu e$ and $\pi\pi$ vertices reconstructed in the data sample agrees with the expectation derived from the prompt background MC sample.


The expected numbers of background events in CR and SR are calculated using a combination of frequentist and Bayesian techniques. The rejection factors of the LAV and ANTI0 veto conditions and the CR and SR selection requirements are defined as the proportion of $e^+e^-$ vertices discarded by the corresponding conditions. The posterior probability distribution function (pdf) of each rejection factor is computed from the prompt background sample, assuming a uniform prior and a beta function likelihood. Pseudo-experiments are generated, sampling independently the number of events in the FV and the rejection factors. The expected numbers of background events are


\begin{align}
\label{eq:finalBkg2}
    \begin{aligned}
        N_{\mathrm{bkg}}^{\mathrm{CR}} &= 
        9.7^{+21.3}_{-7.3}\times 10^{-3} \text{,}~
        N_{\mathrm{bkg}}^{\mathrm{SR}} = 9.4^{+20.6}_{-7.2}\times 10^{-3}~ \\
    \end{aligned}
\end{align}
where the uncertainties are quoted at 68\% confidence level (CL).

\section{STATISTICAL ANALYSIS AND RESULTS}
\label{sec:res}
The signal simulation, the uncertainty and efficiency of the signal selection and the expected \mbox{$\darkphoton$} yield are discussed extensively in Appendix~\mbox{\ref{sec:yield}}. The dominating source of uncertainty in the \mbox{$\darkphoton$} yield is given by the number of primary protons impinging on the TAX and is estimated to be 20\%. Additionally, factors such as the size of the signal MC sample and various reconstruction parameters induce a 2.9\% uncertainty in the signal selection efficiency. After unmasking the CR, no events are observed, in agreement with the 98.3\% probability of observing no counts in the null hypothesis. After unmasking the SR, no events are observed, in agreement with the 98.4\% probability of observing no counts in the null hypothesis. The exclusion limits obtained are derived using the $\mathrm{CL}_{s}$ method~\cite{CLsMethod} on a grid of $\darkphoton$ mass and coupling values. The test statistic is the profile likelihood ratio~\cite{ProfiledLikelihood}:
\begin{equation}
  q=-2\ln\frac{L_{s+b}}{L_{b}}, 
\end{equation}
where
\begin{widetext}
\begin{equation}
    L_{s+b} = \rho(\theta|\tilde{\theta})\frac{e^{-(s(\theta)+b(\theta))}}{N}\prod_{i}(s(\theta) f_{s}(x_i; M_{\darkphoton},\varepsilon) + b(\theta) f_{b}(x_i; M_{\darkphoton},\varepsilon))
\end{equation}
\end{widetext}
is the likelihood of the observed data under the signal-plus-background hypothesis. The product runs over the observed events. The terms $s(\theta)$ and $b(\theta)$ are the numbers of signal and background events in the SR, respectively. The functions $f_{s}$ and $f_{b}$ are the signal and background pdfs of the reconstructed mass of the two leptons, $x_i$. The symbol $\theta$ collectively denotes the nuisance parameters: the number of protons on TAX and the expected number of signal and background events in SR. The functions $\rho(\theta|\tilde{\theta})$, where $\tilde\theta$ contains the default values of the nuisance parameters, are the systematic error pdfs. These are interpreted as posteriors derived from simulations. The specific functional forms of all pdfs are given Appendix~\ref{sec:pdfs}.

\begin{figure*}[ht]
    \includegraphics[width=\columnwidth]{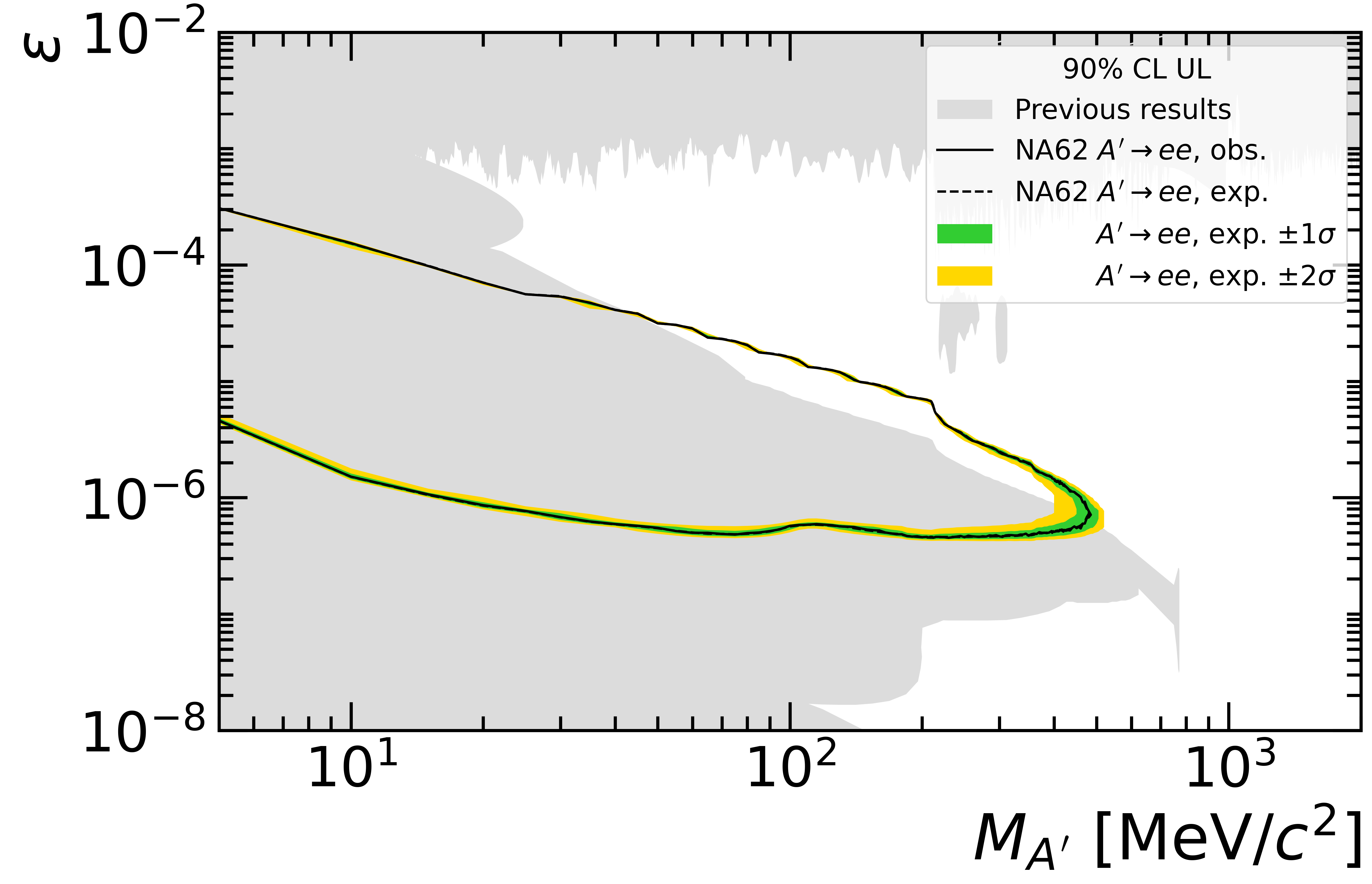}~
    \includegraphics[width=\columnwidth]{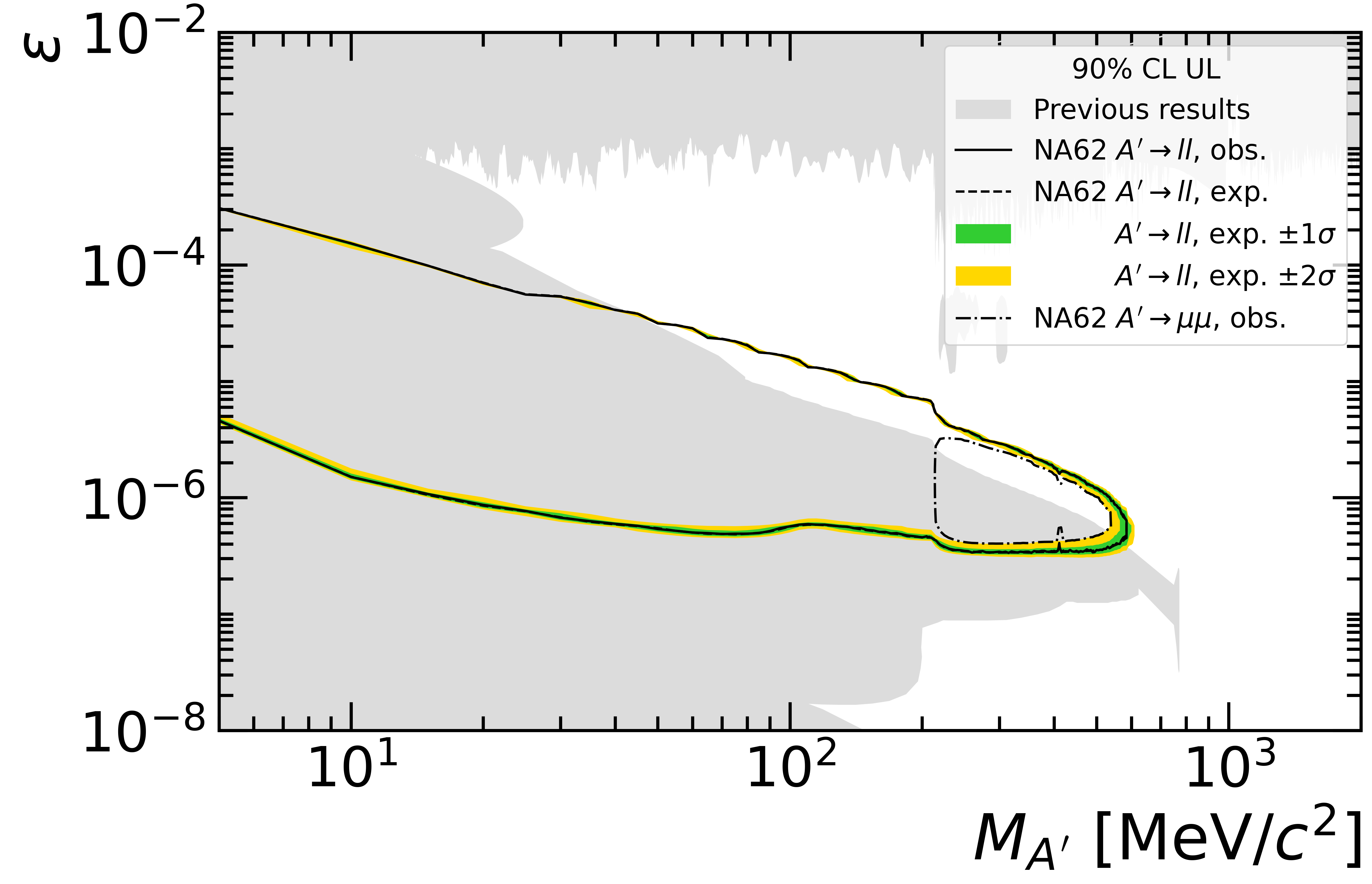}
    \caption{Observed and expected exclusion contours, at 90\% CL, in the plane $\left(M_{\darkphoton},\varepsilon\right)$ for the $\darkphoton\to e^+e^-$ analysis (left) and the combined $\darkphoton\to e^+e^-$ and $\darkphoton\to \mu^+\mu^-$ analyses (right) together with the expected $\pm1\sigma$ (green) and $\pm2\sigma$ (yellow) bands. Previous results, including the recent FASER result~\cite{Abreu_FASER_PLB2024} are shown in grey. The NA62 $\darkphoton\to\mu^+\mu^-$ result~\cite{Dobrich:2023dkm} is shown with a dot-dashed line in the right panel.}
    \label{fig:exclusion_dark_photon}
\end{figure*}

The likelihood of the data under the background-only hypothesis, $L_{b}$, has a similar form, with the signal-related components removed. The exclusion contour is obtained by fitting $q$ to the observation, for each value of $M_{\darkphoton}$ and $\varepsilon$, maximising separately the numerator and denominator with respect to the nuisance parameters. Pseudo-experiments are generated under both signal-only and signal-plus-background hypotheses, using the respective fitted values $\theta$. The same test statistic is computed for each pseudo-experiment. The distributions of $q$ under these hypotheses are used in the $\mathrm{CL}_{s}$ method to decide whether a specific $(M_{\darkphoton},\varepsilon)$ point is excluded or not at a desired confidence level. The observed and expected exclusion contours at 90\% CL, and the expected $\pm1\sigma$ and $\pm2\sigma$ bands, in the $(M_{\darkphoton}, \varepsilon)$ plane are shown in Figure~\ref{fig:exclusion_dark_photon}. Previous results~\cite{Riordan_E141_PRL1987,Davier_Orsay_PLB1989,Blumlein_NuCal_IJMA1992, Andreas_Orsay_PRD2012,Lees_BaBar_PRL2014,Blumlein_PLB2014,Batley_NA48_PLB2015,Banerjee_NA64_2020,Abreu_FASER_PLB2024,Gninenko_NOMAD_PS191_PRD2012,Essig_E137_JHEP2011, Aaij:2017rft, Konaka:1986cb, PhysRevD.82.113008}, adapted from the DarkCast package~\cite{Ilten2018}, and supernova exclusions~\cite{Chang2017} are shown as grey areas.

\begin{figure*}[ht]
    \includegraphics[width=\linewidth]{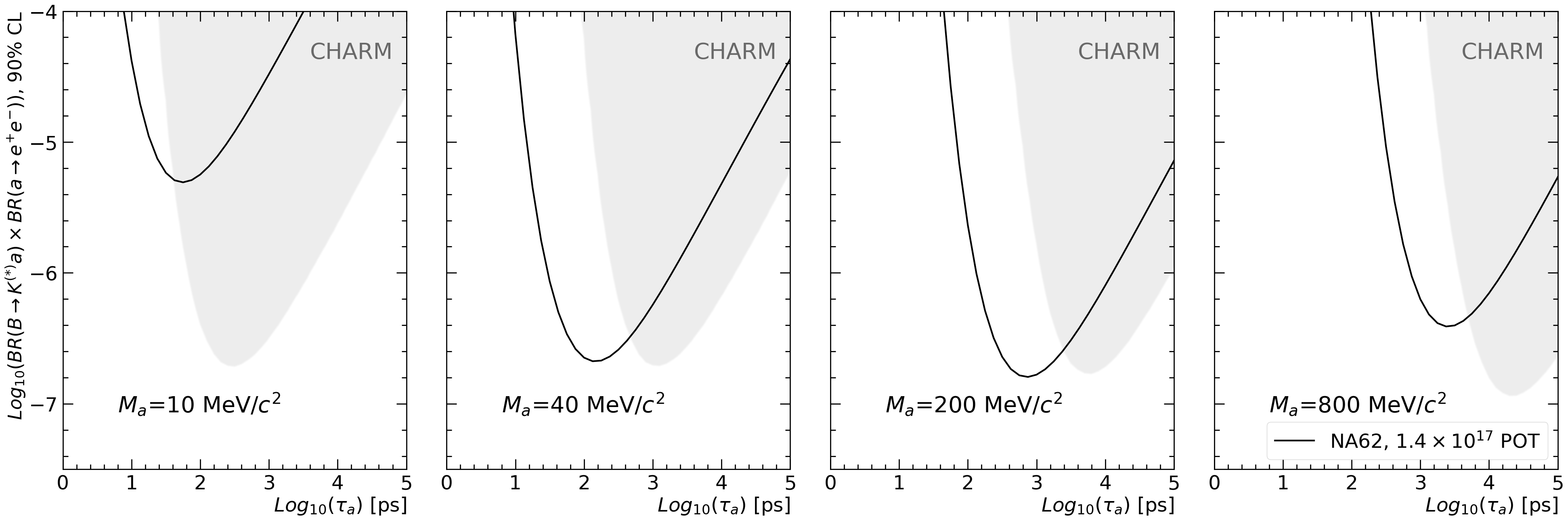}
    \caption{Exclusion region in the plane of the ALP lifetime \mbox{$(\tau_a)$} and the product of branching ratios $BR(B \to K^{(*)}a) \times BR(a \to e^+e^-)$ in the search for an axion-like particle $a$ produced in $B$ meson decays (solid curve). Four values of the ALP mass are considered. The region of the parameter space above the black line is excluded at 90\% CL. The excluded regions by CHARM~\cite{charm} measurements are shown as grey-filled areas. }
    \label{fig:exclusion_alp}
\end{figure*}

The combination of this $\darkphoton\to e^+e^-$ result with the NA62 $\darkphoton\to\mu^+\mu^-$ result~\cite{Dobrich:2023dkm} is performed with the same test statistic but with total likelihoods expressed as products of contributions from the individual $\darkphoton$ decay channels. The number of protons on TAX is common to both channels, therefore its pdf enters only once in the likelihood function. The exclusion regions obtained at 90\% CL are shown in Figure~\ref{fig:exclusion_dark_photon}. 

The interpretation of this $e^+e^-$ result in terms of the emission of ALPs in $b \to s$ transitions is shown in Figure~\ref{fig:exclusion_alp} for a set of ALP mass values. 
A model-independent approach is used, where the ALP lifetime $\tau_a$, the mass $M_a$ and the product $BR(B \to K^{(*)}a) \times BR(a \to e^+ e^-)$ are the free parameters~\cite{Dobrich_PLB2019}. 
Here, $BR(B \to K^{(*)}a)$ stands for the sum of these branching ratios of $B$ meson decays ($B^+$, $B^0$ and their antiparticles) weighted by the corresponding production yields in proton-nucleus interaction obtained from the \texttt{PYTHIA8.2}~\cite{SHORSTRAND_PYTHIA82_CPC2015}  simulation used in~\mbox{\cite{Dobrich_PLB2019}}. 
The same notation was adopted in~\cite{Dobrich:2023dkm}.
The result is found to improve on previous limits in a mass range from 10 to 800~MeV/$c^2$.


\section{Conclusion}
\label{sec:conclusions}

A search for the decay of a dark photon to the $e^+e^-$ final state utilising data taken in beam-dump mode at the NA62 experiment in 2021 is presented. No event is found in the signal region. A statistical combination with a previous search for the $\mu^+\mu^-$ final state by NA62 is performed, extending the previous exclusion limits on dark photons in the mass range from 50 to 600 MeV/$c^2$ and coupling constant range $10^{-6}$ to $4\times 10^{-5}$. The excluded region is compatible with thermal relic density constraints. The interpretation of the $e^+ e^-$ result in terms of the emission of ALPs coupled to the SM fermionic field is also performed, extending the excluded regions in a mass range from 10 to 800~MeV/$c^2$.


\section*{ACKNOWLEDGEMENTS}
\input{acknowrun2021_optC}
\appendix \section{Signal yield and selection efficiency}
\label{sec:yield}


\begin{figure}[htb]
    \includegraphics[width=\columnwidth]{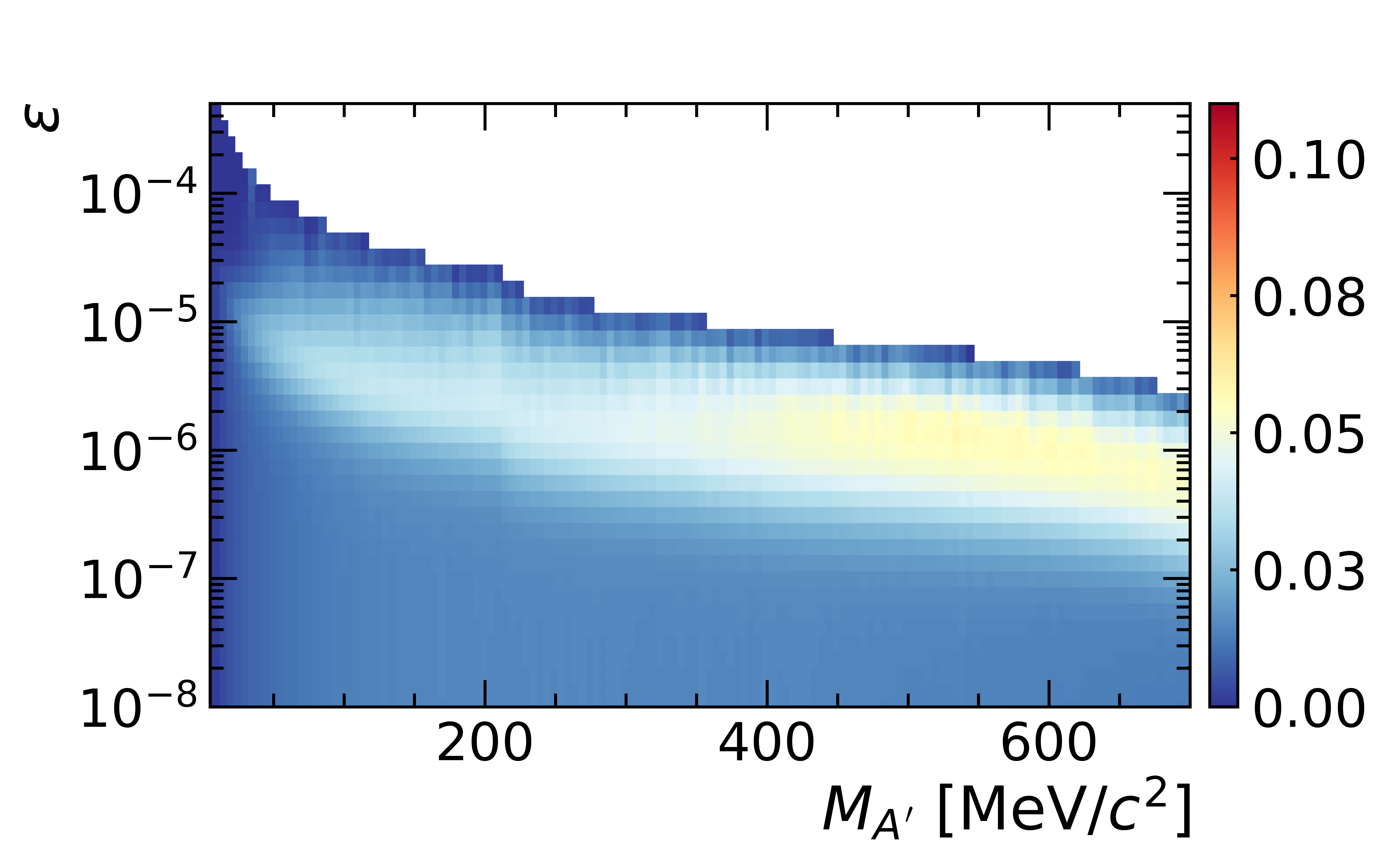}
    \includegraphics[width=\columnwidth]{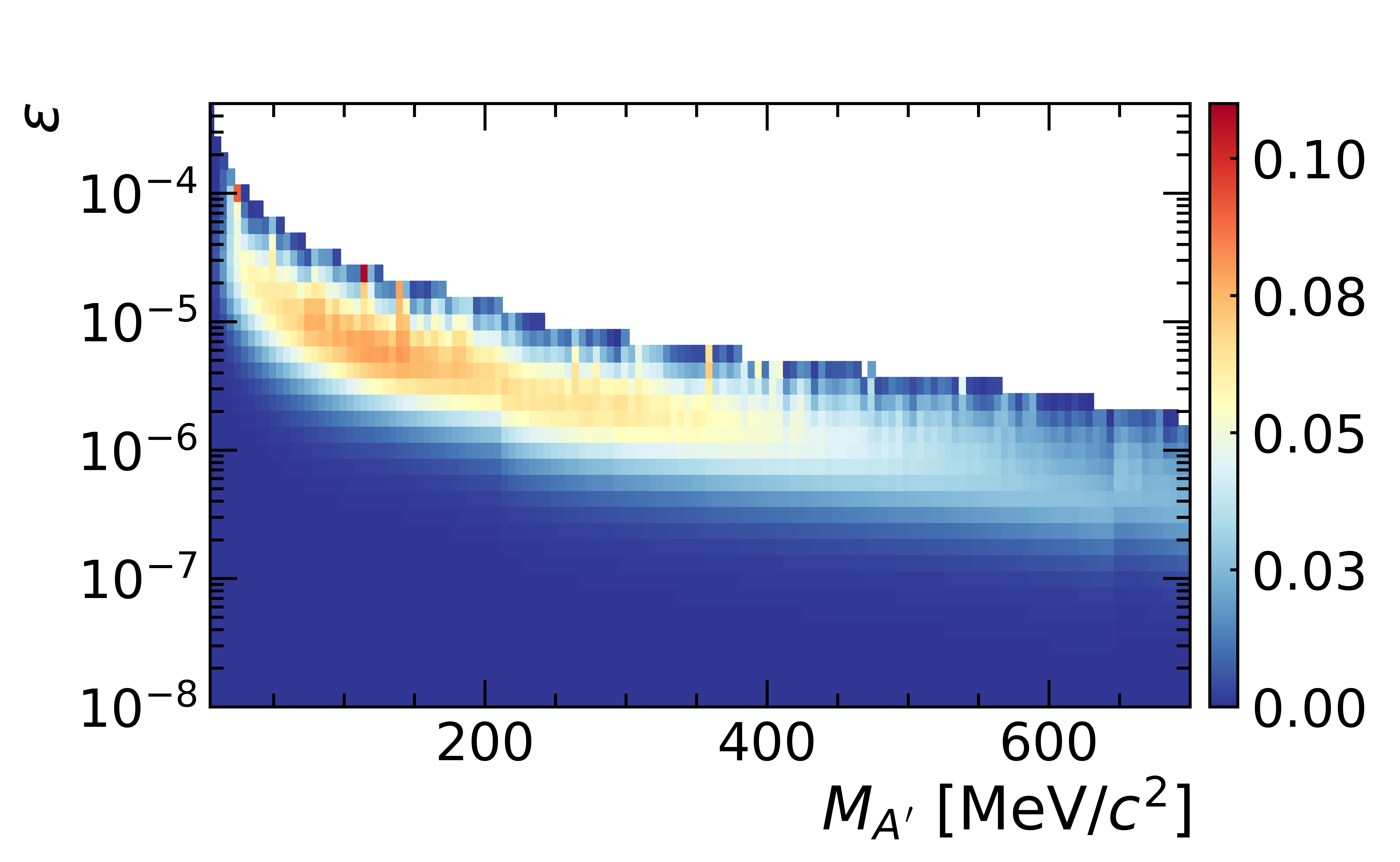}
    \caption{Selection and trigger efficiency (colour scale) for the $\darkphoton\to e^+e^-$ in the plane $(M_{\darkphoton}, \varepsilon)$. The bremsstrahlung (meson-mediated) production mode is shown in the top (bottom) panel.}
    \label{fig:dpee_epsilon}
\end{figure}

The expected $\darkphoton$ yield for given values of the mass and coupling constant is expressed as
\begin{align}
    \begin{aligned}
         N_{\mathrm{exp}} = &N_\mathrm{POT}\times \mathrm{P}(pN\to\darkphoton)\times \mathrm{P}_{\mathrm{D}}\\
         &\times \mathrm{BR}(\darkphoton\to l^+l^-) \times A_{\mathrm{sel}},
    \end{aligned}
\end{align}
where $N_\mathrm{POT}$ is the number of primary protons impinging on the TAX, $\mathrm{P}(pN\to\darkphoton)$ is the $\darkphoton$ production probability per proton, $\mathrm{P}_{\mathrm{D}}$ is the probability for the dark photon to decay within the fiducial volume, $\mathrm{BR}(\darkphoton\to l^+l^-)$ is the branching ratio of the $\darkphoton$ decay to a lepton pair and $A_{\mathrm{sel}}$ is the combined selection and trigger efficiency. 

For every spill, the quantity $N_{\mathrm{POT}}$ is determined by measuring the proton beam flux. This measurement is performed using a titanium-foil secondary-emission monitor positioned at the target location. The uncertainty on $N_{\mathrm{POT}}$ is deduced from the operational experience of these monitors and is estimated to be 20\%. This estimation is confirmed by NA62 kaon decay dataset: the count of selected $K^+\to\pi^+\pi^+\pi^-$ decays matches the expected number based on the measured proton flux within 20\%. 

The uncertainty on the $\darkphoton$ yield in the case of production by bremsstrahlung is given mainly by the uncertainty on the $pp$ scattering cross-section, which is 1\%, as estimated from the available data~\cite{Workman_PDG_PTEP2022}. The branching ratio of neutral meson decays with $\darkphoton$ in the final state is evaluated as in~\cite{Blumlein_PLB2011}. The yield of various neutral mesons in $pp$ interactions at $\sqrt{s}=28$~GeV contributing to the $\darkphoton$ yield is evaluated with \verb|PYTHIA8.2|~\cite{SHORSTRAND_PYTHIA82_CPC2015}. The meson production cross-sections have been validated against data and their uncertainties are estimated
to be at the 20\% level~\cite{Dobrich_JHEP2019}. The $\darkphoton$ probability to decay within the fiducial volume is computed for each considered point in the parameter space. The $\darkphoton \to e^+e^-$ decay branching ratio is evaluated according to~\mbox{\cite{Batell_PRD2009}}.

Monte Carlo simulations of $\darkphoton$ production and decay are used to evaluate the combined selection and trigger efficiency at selected values in the $(M_{\darkphoton}, \varepsilon)$ plane. The $\darkphoton$ mass is varied from 5~MeV/$c^2$ to 700~MeV/$c^2$, and the $\darkphoton$ is forced to decay within the fiducial
volume using a uniform decay distribution. Each event is weighted by the decay probability $\textrm{P}_\textrm{D}$, which depends on the coupling constant $\varepsilon$. The efficiencies are shown in Figure~\ref{fig:dpee_epsilon}. 

\begin{figure}[htb]
    \includegraphics[width=\columnwidth]{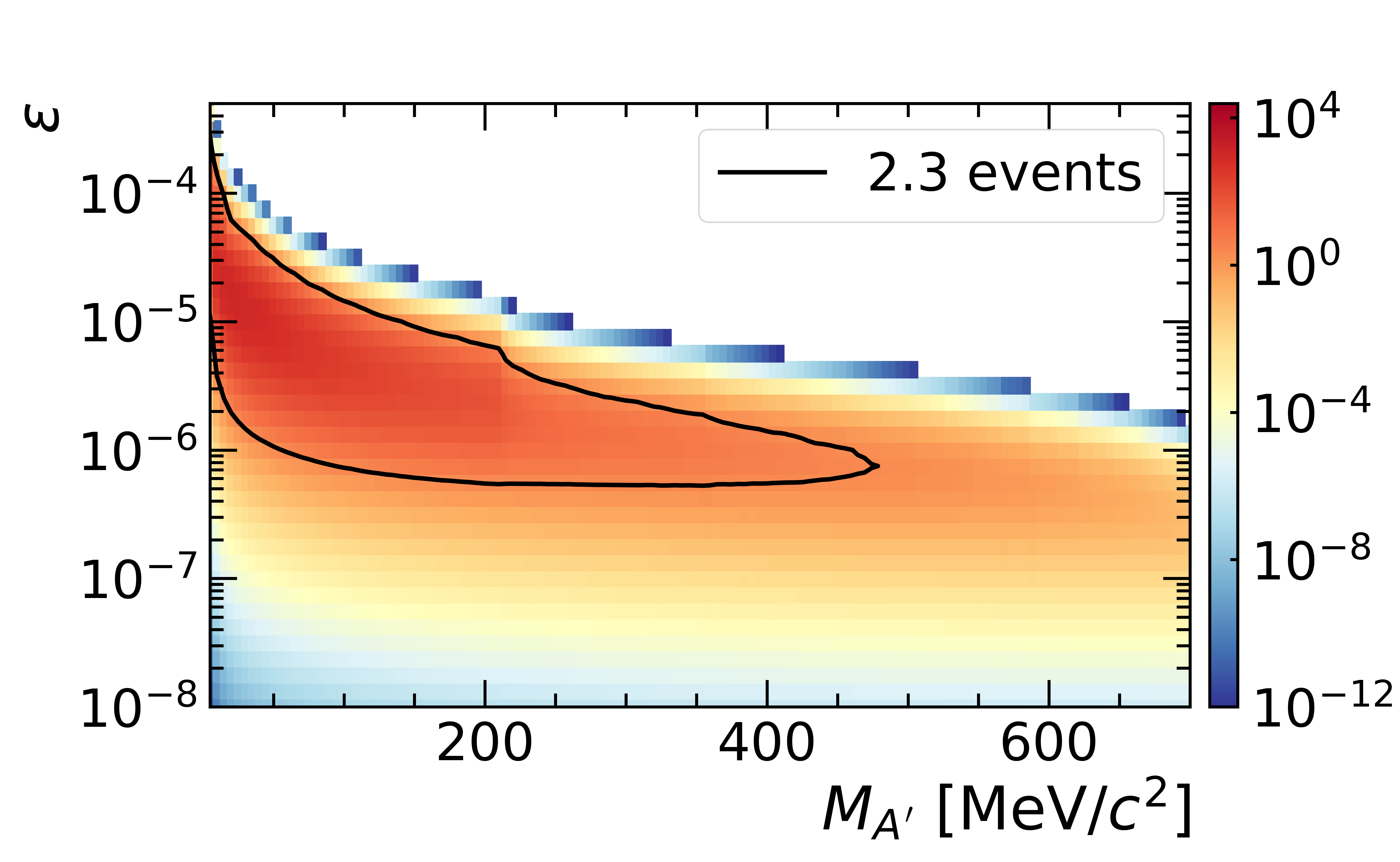}
    \includegraphics[width=\columnwidth]{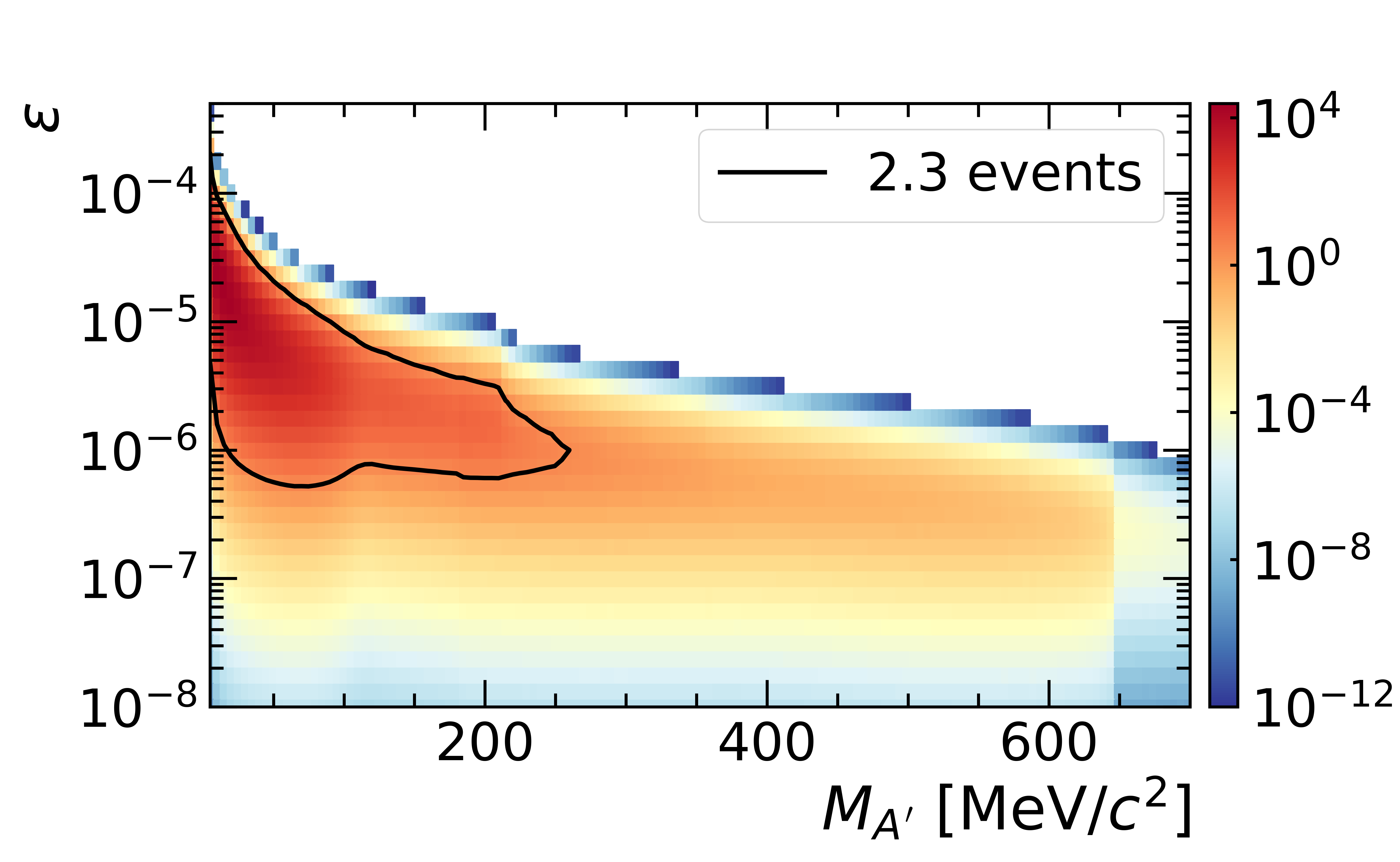}
    \caption{Expected number of events in SR (colour scale) for the $\darkphoton\to e^+e^-$ decay in the plane $(M_{\darkphoton}, \varepsilon)$. The bremsstrahlung (meson-mediated) production mode is shown in the top (bottom) panel. The black contour corresponds to 2.3 events.}
    \label{fig:dpeeyield}
\end{figure}

The expected $\darkphoton\to e^+e^-$ yield is shown in Figure~\ref{fig:dpeeyield}. For comparison, Figure~11 of~\cite{Dobrich:2023dkm} displays the same quantities for the $\darkphoton\to \mu^+\mu^-$ channel. The bremsstrahlung production mode is dominant, therefore the total uncertainty on the yield is dominated by the uncertainty on the number of primary protons impinging on the TAX.

\begin{figure}[htb]
    \centering
    \includegraphics[width=\columnwidth]{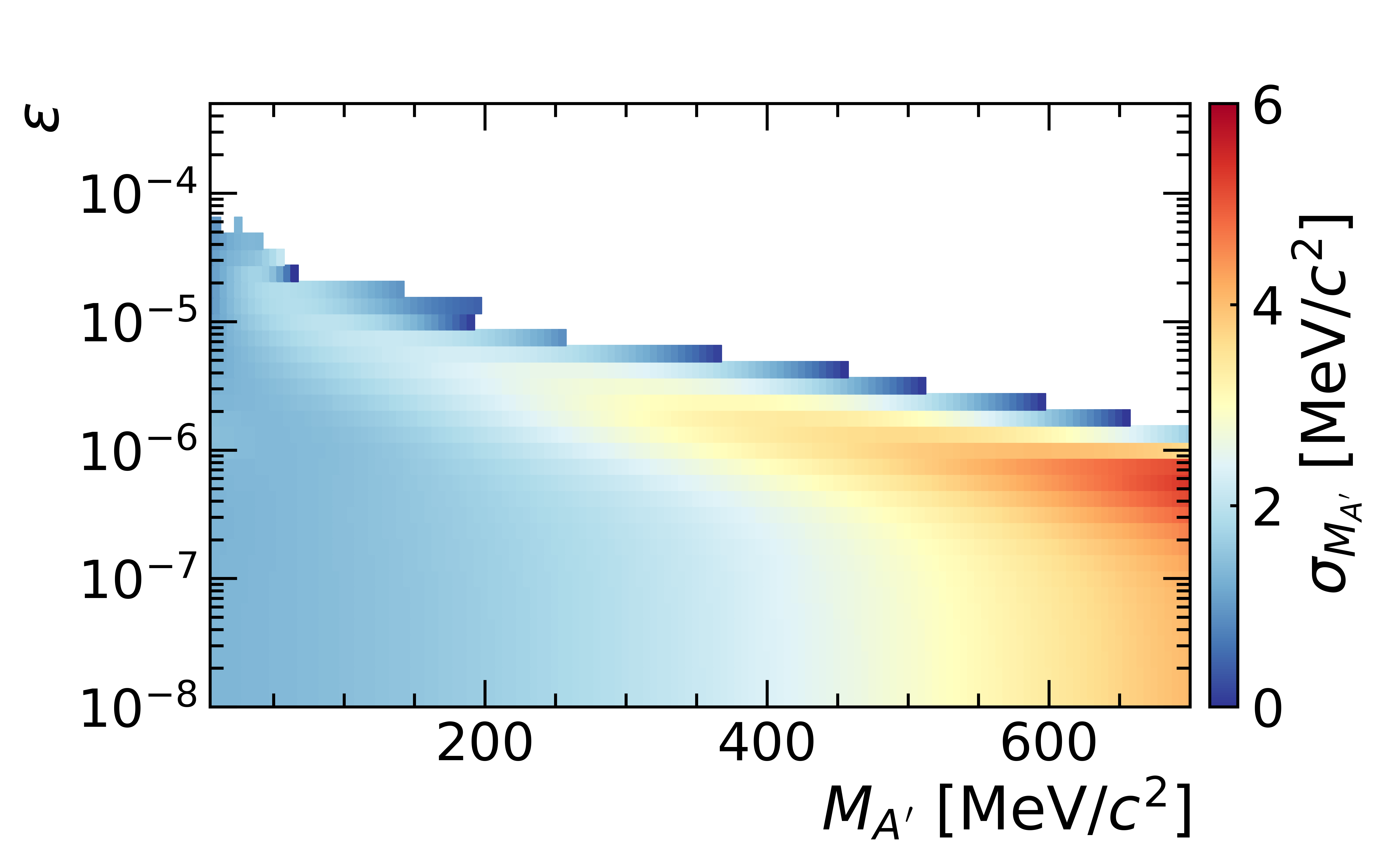}
    \includegraphics[width=\columnwidth]{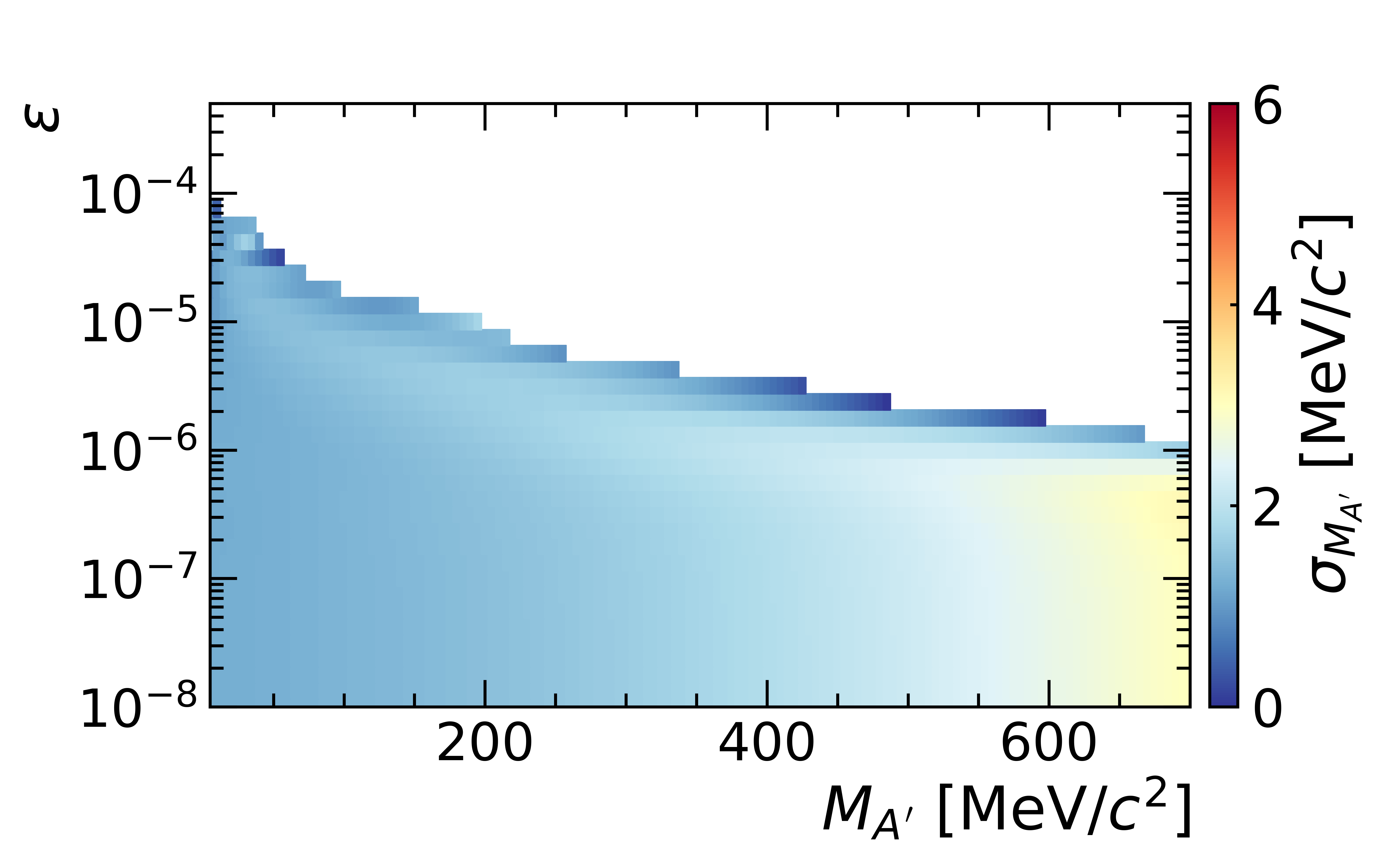}
    \caption{Mass resolution (colour scale) in the plane $(M_{\darkphoton},\varepsilon)$. The top (bottom) panel refers to bremsstrahlung (meson-mediated) production.}
    \label{fig:sigma_ma}
\end{figure}

The $\darkphoton$ mass resolution, $\sigma_{M_{\darkphoton}}$, depends on both $M_{\darkphoton}$ and $\varepsilon$, and differs by production and decay channel. Figure~\ref{fig:sigma_ma} displays this quantity for the $\darkphoton\to e^+e^-$ signal and Figure 12 of~\cite{Dobrich:2023dkm} for $\darkphoton \to \mu^+\mu^-$.

Table~\ref{tab:signal_sel_uncert} summarises the uncertainties affecting the signal selection efficiency. The contribution of each source is assessed via a combination of control samples and simulation. The simulation contribution represents a typical value, as it varies with $M_{\darkphoton}$ and $\varepsilon$. 

\begin{table}[htb]
    \begin{ruledtabular}    
    \begin{tabular}{|c|c|}
    Source  & Uncertainty \\
    \hline
    Simulation & 2.1\% \\
    CHOD association & 0.6\% \\
    PID & 1.1\% \\
    Spectrometer quality and resolution & 1.5\% \\
    Trigger & 0.5\% \\
    LAV random veto & 0.1\% \\
    ANTI0 random veto & 0.1\% \\
    \hline
    Total & 2.9\% \\
    \end{tabular}
    \end{ruledtabular}
    \caption{Summary of uncertainties to the signal selection efficiency\label{tab:signal_sel_uncert}}
    \label{tab:my_label}
\end{table}

\section{Probability distribution functions used in the likelihoods}
\label{sec:pdfs}
All pdfs used in the likelihoods are defined using the \verb|ROOFIT| package~\cite{Verkerke_ROOFIT}. In each point of the $(M_{\darkphoton}, \varepsilon)$ grid used in the determination of the exclusion region, the signal pdfs of the reconstructed di-lepton invariant mass are normal distributions, centred in $M_{\darkphoton}$ and with standard deviation equal to $\sigma_{M_{\darkphoton}}$ (Figure~\ref{fig:sigma_ma} for the $\darkphoton\to e^+e^-$ channel and Figure 12 of~\cite{Dobrich:2023dkm} for $\darkphoton\to \mu^+\mu^-$). The background pdf for the di-muon channel is defined as a linear combination of elements in the Bernstein polynomial basis of degree five, with coefficients
\begin{align}
    \begin{aligned}
        c = (& 1.886\times 10^{-2}, 1.945\times 10^{-1}, 9.183\times 10^{-1},\\ 
        &2.769\times 10^{-1}, 1.196\times 10^{-1}, 6.845\times 10^{-2}),
    \end{aligned}
\end{align}
and by a Landau function with location and shape parameters 14.86~MeV/$c^2$ and 3.48~MeV/$c^2$, respectively, for the $e^+e^-$ channel.

The pdf of the number of protons on TAX is a log-normal distribution with median $1.4\times 10^{17}$ and shape parameter $1.2$. The background yields are modelled by log-normal distributions as well. The median and shape parameters for the $\mu^+\mu^-$ channel are 0.016 and 1.125, respectively. For the $e^+ e^-$ channel, the parameters are 0.0094 and 4.0.

\bigskip

\bibliography{bibliography}

\bigskip
\clearpage
\onecolumngrid
\input{run2-bdeevf}

\end{document}

%% file: acknowrun2021_optC.tex
It is a pleasure to express our appreciation to the staff of the CERN laboratory and the technical
staff of the participating laboratories and universities for their efforts in the operation of the
experiment and data processing.

The cost of the experiment and its auxiliary systems was supported by the funding agencies of 
the Collaboration Institutes. We are particularly indebted to: 
F.R.S.-FNRS (Fonds de la Recherche Scientifique - FNRS), under Grants No. 4.4512.10, 1.B.258.20, Belgium;
CECI (Consortium des Equipements de Calcul Intensif), funded by the Fonds de la Recherche Scientifique de Belgique (F.R.S.-FNRS) under Grant No. 2.5020.11 and by the Walloon Region, Belgium;
NSERC (Natural Sciences and Engineering Research Council), funding SAPPJ-2018-0017,  Canada;
MEYS (Ministry of Education, Youth and Sports) funding LM 2018104, Czech Republic;
BMBF (Bundesministerium f\"{u}r Bildung und Forschung) contracts 05H12UM5, 05H15UMCNA and 05H18UMCNA, Germany;
INFN  (Istituto Nazionale di Fisica Nucleare),  Italy;
MIUR (Ministero dell'Istruzione, dell'Universit\`a e della Ricerca),  Italy;
CONACyT  (Consejo Nacional de Ciencia y Tecnolog\'{i}a),  Mexico;
IFA (Institute of Atomic Physics) Romanian 
CERN-RO Nr. 10/10.03.2020
and Nucleus Programme PN 19 06 01 04,  Romania;
MESRS  (Ministry of Education, Science, Research and Sport), Slovakia; 
CERN (European Organization for Nuclear Research), Switzerland; 
STFC (Science and Technology Facilities Council), United Kingdom;
NSF (National Science Foundation) Award Numbers 1506088 and 1806430,  U.S.A.;
ERC (European Research Council)  ``UniversaLepto'' advanced grant 268062, ``KaonLepton'' starting grant 336581, Europe.

Individuals have received support from:
Charles University (Research Center UNCE/SCI/013, grant PRIMUS 23/SCI/025), Czech Republic;
Czech Science Foundation (grant 23-06770S);
Ministero dell'Istruzione, dell'Universit\`a e della Ricerca (MIUR  ``Futuro in ricerca 2012''  grant RBFR12JF2Z, Project GAP), Italy;
the Royal Society  (grants UF100308, UF0758946), United Kingdom;
STFC (Rutherford fellowships ST/J00412X/1, ST/M005798/1), United Kingdom;
ERC (grants 268062,  336581 and  starting grant 802836 ``AxScale'');
EU Horizon 2020 (Marie Sk\l{}odowska-Curie grants 701386, 754496, 842407, 893101, 101023808).

%% file: run2-bdeevf.tex
\newcommand{\orcimg}{\raisebox{-0.3\height}{\includegraphics[height=\fontcharht\font`A]{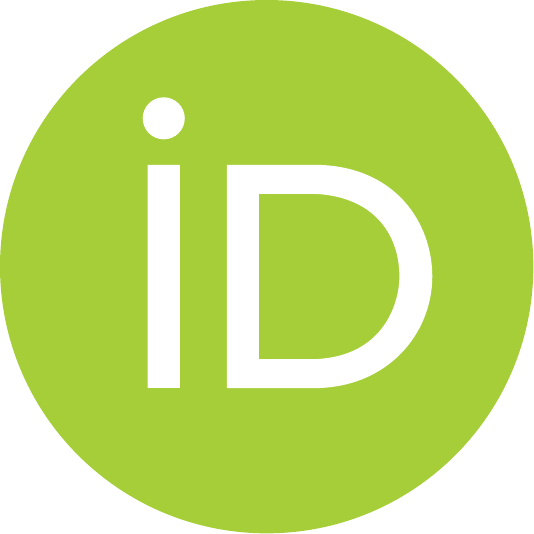}}}
\newcommand{\orcid}[1]{\href{https://orcid.org/#1}{\orcimg}}

\centerline{\bf The NA62 Collaboration} 
\vspace{1.5cm}
%
%

\begin{raggedright}
\noindent
{\bf Universit\'e Catholique de Louvain, Louvain-La-Neuve, Belgium}\\
 E.~Cortina Gil\orcid{0000-0001-9627-699X},
 J.~Jerhot$\,${\footnotemark[1]}\orcid{0000-0002-3236-1471},
 A.~Kleimenova$\,$\renewcommand{\thefootnote}{\fnsymbol{footnote}}\footnotemark[1]\renewcommand{\thefootnote}{\arabic{footnote}}$^,$$\,${\footnotemark[2]}\orcid{0000-0002-9129-4985},
 N.~Lurkin\orcid{0000-0002-9440-5927},
 M.~Zamkovsky$\,${\footnotemark[3]}\orcid{0000-0002-5067-4789}
\vspace{0.5cm}

{\bf TRIUMF, Vancouver, British Columbia, Canada}\\
 T.~Numao\orcid{0000-0001-5232-6190},
 B.~Velghe\orcid{0000-0002-0797-8381},
 V. W. S.~Wong\orcid{0000-0001-5975-8164}
\vspace{0.5cm}

{\bf University of British Columbia, Vancouver, British Columbia, Canada}\\
 D.~Bryman$\,${\footnotemark[4]}\orcid{0000-0002-9691-0775}
\vspace{0.5cm}

{\bf Charles University, Prague, Czech Republic}\\
 Z.~Hives\orcid{0000-0002-5025-993X},
 T.~Husek$\,${\footnotemark[5]}\orcid{0000-0002-7208-9150},
 K.~Kampf\orcid{0000-0003-1096-667X},
 M.~Koval\orcid{0000-0002-6027-317X}
\vspace{0.5cm}

{\bf Aix Marseille University, CNRS/IN2P3, CPPM, Marseille, France}\\
 B.~De Martino\orcid{0000-0003-2028-9326},
 M.~Perrin-Terrin\orcid{0000-0002-3568-1956}
\vspace{0.5cm}

{\bf Max-Planck-Institut f\"ur Physik (Werner-Heisenberg-Institut), Garching, Germany}\\
 B.~D\"obrich\orcid{0000-0002-6008-8601},
 S.~Lezki\orcid{0000-0002-6909-774X}
\vspace{0.5cm}

{\bf Institut f\"ur Physik and PRISMA Cluster of Excellence, Universit\"at Mainz, Mainz, Germany}\\
 A.~T.~Akmete\orcid{0000-0002-5580-5477},
 R.~Aliberti$\,${\footnotemark[6]}\orcid{0000-0003-3500-4012},
 L.~Di Lella\orcid{0000-0003-3697-1098},
 N.~Doble\orcid{0000-0002-0174-5608},
 L.~Peruzzo\orcid{0000-0002-4752-6160}, 
 S.~Schuchmann\orcid{0000-0002-8088-4226},
 H.~Wahl\orcid{0000-0003-0354-2465},
 R.~Wanke\orcid{0000-0002-3636-360X}
\vspace{0.5cm}

{\bf Dipartimento di Fisica e Scienze della Terra dell'Universit\`a e INFN, Sezione di Ferrara, Ferrara, Italy}\\
 P.~Dalpiaz,
 I.~Neri\orcid{0000-0002-9669-1058},
 F.~Petrucci\orcid{0000-0002-7220-6919},
 M.~Soldani\orcid{0000-0003-4902-943X}
\vspace{0.5cm}

{\bf INFN, Sezione di Ferrara, Ferrara, Italy}\\
 L.~Bandiera\orcid{0000-0002-5537-9674},
 A.~Cotta Ramusino\orcid{0000-0003-1727-2478},
 A.~Gianoli\orcid{0000-0002-2456-8667},
 M.~Romagnoni\orcid{0000-0002-2775-6903},
 A.~Sytov\orcid{0000-0001-8789-2440}
\vspace{0.5cm}

{\bf Dipartimento di Fisica e Astronomia dell'Universit\`a e INFN, Sezione di Firenze, Sesto Fiorentino, Italy}\\
 M.~Lenti\orcid{0000-0002-2765-3955},
 P.~Lo Chiatto\orcid{0000-0002-4177-557X},
 R.~Marchevski$\,${\footnotemark[2]}\orcid{0000-0003-3410-0918},
 I.~Panichi\orcid{0000-0001-7749-7914},
 G.~Ruggiero\orcid{0000-0001-6605-4739}
\vspace{0.5cm}

{\bf INFN, Sezione di Firenze, Sesto Fiorentino, Italy}\\
 A.~Bizzeti$\,${\footnotemark[7]}\orcid{0000-0001-5729-5530},
 F.~Bucci\orcid{0000-0003-1726-3838}
\vspace{0.5cm}

{\bf Laboratori Nazionali di Frascati, Frascati, Italy}\\
 A.~Antonelli\orcid{0000-0001-7671-7890},
 V.~Kozhuharov$\,${\footnotemark[8]}\orcid{0000-0002-0669-7799},
 G.~Lanfranchi\orcid{0000-0002-9467-8001},
 S.~Martellotti\orcid{0000-0002-4363-7816},
 M.~Moulson\orcid{0000-0002-3951-4389}, 
 T.~Spadaro\orcid{0000-0002-7101-2389},
 G.~Tinti\orcid{0000-0003-1364-844X}
\vspace{0.5cm}

{\bf Dipartimento di Fisica ``Ettore Pancini'' e INFN, Sezione di Napoli, Napoli, Italy}\\
 F.~Ambrosino\orcid{0000-0001-5577-1820},
 M.~D'Errico\orcid{0000-0001-5326-1106},
 R.~Fiorenza$\,${\footnotemark[9]}\orcid{0000-0003-4965-7073},
 R.~Giordano\orcid{0000-0002-5496-7247},
 P.~Massarotti\orcid{0000-0002-9335-9690}, 
 M.~Mirra\orcid{0000-0002-1190-2961},
 M.~Napolitano\orcid{0000-0003-1074-9552},
 I.~Rosa\orcid{0009-0002-7564-182},
 G.~Saracino\orcid{0000-0002-0714-5777}
\vspace{0.5cm}

{\bf Dipartimento di Fisica e Geologia dell'Universit\`a e INFN, Sezione di Perugia, Perugia, Italy}\\
 G.~Anzivino\orcid{0000-0002-5967-0952}
\vspace{0.5cm}

{\bf INFN, Sezione di Perugia, Perugia, Italy}\\
 F.~Brizioli$\,${\footnotemark[3]}\orcid{0000-0002-2047-441X},
 P.~Cenci\orcid{0000-0001-6149-2676},
 V.~Duk\orcid{0000-0001-6440-0087},
 R.~Lollini\orcid{0000-0003-3898-7464},
 P.~Lubrano\orcid{0000-0003-0221-4806}, 
 M.~Pepe\orcid{0000-0001-5624-4010},
 M.~Piccini\orcid{0000-0001-8659-4409}
\vspace{0.5cm}
\newpage
{\bf Dipartimento di Fisica dell'Universit\`a e INFN, Sezione di Pisa, Pisa, Italy}\\
 F.~Costantini\orcid{0000-0002-2974-0067},
 M.~Giorgi\orcid{0000-0001-9571-6260},
 S.~Giudici\orcid{0000-0003-3423-7981},
 G.~Lamanna\orcid{0000-0001-7452-8498},
 E.~Lari\orcid{0000-0003-3303-0524}, 
 E.~Pedreschi\orcid{0000-0001-7631-3933},
 J.~Pinzino\orcid{0000-0002-7418-0636},
 M.~Sozzi\orcid{0000-0002-2923-1465}
\vspace{0.5cm}

{\bf INFN, Sezione di Pisa, Pisa, Italy}\\
 R.~Fantechi\orcid{0000-0002-6243-5726},
 F.~Spinella\orcid{0000-0002-9607-7920}
\vspace{0.5cm}

{\bf Scuola Normale Superiore e INFN, Sezione di Pisa, Pisa, Italy}\\
 I.~Mannelli\orcid{0000-0003-0445-7422}
\vspace{0.5cm}

{\bf Dipartimento di Fisica, Sapienza Universit\`a di Roma e INFN, Sezione di Roma I, Roma, Italy}\\
 M.~Raggi\orcid{0000-0002-7448-9481}
\vspace{0.5cm}

{\bf INFN, Sezione di Roma I, Roma, Italy}\\
 A.~Biagioni\orcid{0000-0001-5820-1209},
 P.~Cretaro\orcid{0000-0002-2229-149X},
 O.~Frezza\orcid{0000-0001-8277-1877},
 A.~Lonardo\orcid{0000-0002-5909-6508},
 M.~Turisini\orcid{0000-0002-5422-1891},
 P.~Vicini\orcid{0000-0002-4379-4563}
\vspace{0.5cm}

{\bf INFN, Sezione di Roma Tor Vergata, Roma, Italy}\\
 R.~Ammendola\orcid{0000-0003-4501-3289},
 V.~Bonaiuto$\,${\footnotemark[10]}\orcid{0000-0002-2328-4793},
 A.~Fucci,
 A.~Salamon\orcid{0000-0002-8438-8983},
 F.~Sargeni$\,${\footnotemark[11]}\orcid{0000-0002-0131-236X}
\vspace{0.5cm}

{\bf Dipartimento di Fisica dell'Universit\`a e INFN, Sezione di Torino, Torino, Italy}\\
 R.~Arcidiacono$\,${\footnotemark[12]}\orcid{0000-0001-5904-142X},
 B.~Bloch-Devaux\orcid{0000-0002-2463-1232},
 E.~Menichetti\orcid{0000-0001-7143-8200},
 E.~Migliore\orcid{0000-0002-2271-5192}
\vspace{0.5cm}

{\bf INFN, Sezione di Torino, Torino, Italy}\\
 C.~Biino\orcid{0000-0002-1397-7246},
 A.~Filippi\orcid{0000-0003-4715-8748},
 F.~Marchetto\orcid{0000-0002-5623-8494},
 D.~Soldi\orcid{0000-0001-9059-4831}
\vspace{0.5cm}

{\bf Instituto de F\'isica, Universidad Aut\'onoma de San Luis Potos\'i, San Luis Potos\'i, Mexico}\\
 A.~Briano Olvera\orcid{0000-0001-6121-3905},
 J.~Engelfried\orcid{0000-0001-5478-0602},
 N.~Estrada-Tristan$\,${\footnotemark[13]}\orcid{0000-0003-2977-9380},
 R.~Piandani\orcid{0000-0003-2226-8924},
 M. A.~Reyes Santos$\,${\footnotemark[13]}\orcid{0000-0003-1347-2579},
 K. A.~Rodriguez Rivera\orcid{0000-0001-5723-9176}
\vspace{0.5cm}

{\bf Horia Hulubei National Institute for R\&D in Physics and Nuclear Engineering, Bucharest-Magurele, Romania}\\
 P.~Boboc\orcid{0000-0001-5532-4887},
 A. M.~Bragadireanu,
 S. A.~Ghinescu$\,$\renewcommand{\thefootnote}{\fnsymbol{footnote}}\footnotemark[1]\renewcommand{\thefootnote}{\arabic{footnote}}\orcid{0000-0003-3716-9857},
 O. E.~Hutanu
\vspace{0.5cm}

{\bf Faculty of Mathematics, Physics and Informatics, Comenius University, Bratislava, Slovakia}\\
 T.~Blazek\orcid{0000-0002-2645-0283},
 V.~Cerny\orcid{0000-0003-1998-3441},
 Z.~Kucerova$\,${\footnotemark[3]}\orcid{0000-0001-8906-3902},
 R.~Volpe$\,${\footnotemark[14]}\orcid{0000-0003-1782-2978}
\vspace{0.5cm}

{\bf CERN, European Organization for Nuclear Research, Geneva, Switzerland}\\
 J.~Bernhard\orcid{0000-0001-9256-971X},
 L.~Bician$\,${\footnotemark[15]}\orcid{0000-0001-9318-0116},
 M.~Boretto\orcid{0000-0001-5012-4480},
 A.~Ceccucci\orcid{0000-0002-9506-866X},
 M.~Ceoletta\orcid{0000-0002-2532-0217}, 
 M.~Corvino\orcid{0000-0002-2401-412X},
 H.~Danielsson\orcid{0000-0002-1016-5576},
 F.~Duval,
 L.~Federici\orcid{0000-0002-3401-9522},
 E.~Gamberini\orcid{0000-0002-6040-4985}, 
 R.~Guida,
 E.~B.~Holzer\orcid{0000-0003-2622-6844},
 B.~Jenninger,
 G.~Lehmann Miotto\orcid{0000-0001-9045-7853},
 P.~Lichard\orcid{0000-0003-2223-9373}, 
 K.~Massri\orcid{0000-0001-7533-6295},
 E.~Minucci$\,${\footnotemark[16]}\orcid{0000-0002-3972-6824},
 M.~Noy,
 V.~Ryjov,
 J.~Swallow$\,${\footnotemark[17]}\orcid{0000-0002-1521-0911}
\vspace{0.5cm}

{\bf School of Physics and Astronomy, University of Birmingham, Birmingham, United Kingdom}\\
 J. R.~Fry\orcid{0000-0002-3680-361X},
 F.~Gonnella\orcid{0000-0003-0885-1654},
 E.~Goudzovski\orcid{0000-0001-9398-4237},
 J.~Henshaw\orcid{0000-0001-7059-421X},
 C.~Kenworthy\orcid{0009-0002-8815-0048}, 
 C.~Lazzeroni\orcid{0000-0003-4074-4787},
 C.~Parkinson\orcid{0000-0003-0344-7361},
 A.~Romano\orcid{0000-0003-1779-9122},
 J.~Sanders\orcid{0000-0003-1014-094X},
 A.~Sergi$\,${\footnotemark[18]}\orcid{0000-0001-9495-6115}, 
 A.~Shaikhiev$\,${\footnotemark[19]}\orcid{0000-0003-2921-8743},
 A.~Tomczak\orcid{0000-0001-5635-3567}
\vspace{0.5cm}

{\bf School of Physics, University of Bristol, Bristol, United Kingdom}\\
 H.~Heath\orcid{0000-0001-6576-9740}
\vspace{0.5cm}

{\bf School of Physics and Astronomy, University of Glasgow, Glasgow, United Kingdom}\\
 D.~Britton\orcid{0000-0001-9998-4342},
 A.~Norton\orcid{0000-0001-5959-5879},
 D.~Protopopescu\orcid{0000-0002-3964-3930}
\vspace{0.5cm}

{\bf Physics Department, University of Lancaster, Lancaster, United Kingdom}\\
 J. B.~Dainton,
 L.~Gatignon\orcid{0000-0001-6439-2945},
 R. W. L.~Jones\orcid{0000-0002-6427-3513}
\vspace{0.5cm}
\newpage
{\bf Physics and Astronomy Department, George Mason University, Fairfax, Virginia, USA}\\
 P.~Cooper,
 D.~Coward$\,${\footnotemark[20]}\orcid{0000-0001-7588-1779},
 P.~Rubin\orcid{0000-0001-6678-4985}
\vspace{0.5cm}

{\bf Authors affiliated with an Institute or an international laboratory covered by a cooperation agreement with CERN}\\
 A.~Baeva,
 D.~Baigarashev$\,${\footnotemark[21]}\orcid{0000-0001-6101-317X},
 D.~Emelyanov,
 T.~Enik\orcid{0000-0002-2761-9730},
 V.~Falaleev$\,${\footnotemark[14]}\orcid{0000-0003-3150-2196}, 
 S.~Fedotov,
 K.~Gorshanov\orcid{0000-0001-7912-5962},
 E.~Gushchin\orcid{0000-0001-8857-1665},
 V.~Kekelidze\orcid{0000-0001-8122-5065},
 D.~Kereibay, 
 S.~Kholodenko$\,${\footnotemark[22]}\orcid{0000-0002-0260-6570},
 A.~Khotyantsev,
 A.~Korotkova,
 Y.~Kudenko\orcid{0000-0003-3204-9426},
 V.~Kurochka, 
 V.~Kurshetsov\orcid{0000-0003-0174-7336},
 L.~Litov$\,${\footnotemark[8]}\orcid{0000-0002-8511-6883},
 D.~Madigozhin\orcid{0000-0001-8524-3455},
 A.~Mefodev,
 M.~Misheva$\,${\footnotemark[23]}, 
 N.~Molokanova,
 V.~Obraztsov\orcid{0000-0002-0994-3641},
 A.~Okhotnikov\orcid{0000-0003-1404-3522},
 I.~Polenkevich,
 Yu.~Potrebenikov\orcid{0000-0003-1437-4129}, 
 A.~Sadovskiy\orcid{0000-0002-4448-6845},
 S.~Shkarovskiy,
 V.~Sugonyaev\orcid{0000-0003-4449-9993},
 O.~Yushchenko\orcid{0000-0003-4236-5115}
\vspace{0.5cm}

\end{raggedright}

%
%

\setcounter{footnote}{0}
\newlength{\basefootnotesep}
\setlength{\basefootnotesep}{\footnotesep}

\renewcommand{\thefootnote}{\fnsymbol{footnote}}
\noindent
$^{\footnotemark[1]}${Corresponding authors: S.~Ghinescu, A.~Kleimenova,\\
email:  stefan.ghinescu@cern.ch, alina.kleimenova@cern.ch}\\
\renewcommand{\thefootnote}{\arabic{footnote}}
$^{1}${Present address: Max-Planck-Institut f\"ur Physik (Werner-Heisenberg-Institut), D-85748 Garching, Germany} \\
$^{2}${Present address: Ecole Polytechnique F\'ed\'erale Lausanne, CH-1015 Lausanne, Switzerland} \\
$^{3}${Present address: CERN, European Organization for Nuclear Research, CH-1211 Geneva 23, Switzerland} \\
$^{4}${Also at TRIUMF, Vancouver, British Columbia, V6T 2A3, Canada} \\
$^{5}${Also at School of Physics and Astronomy, University of Birmingham, Birmingham, B15 2TT, UK} \\
$^{6}${Present address: Institut f\"ur Kernphysik and Helmholtz Institute Mainz, Universit\"at Mainz, Mainz, D-55099, Germany} \\
$^{7}${Also at Dipartimento di Scienze Fisiche, Informatiche e Matematiche, Universit\`a di Modena e Reggio Emilia, I-41125 Modena, Italy} \\
$^{8}${Also at Faculty of Physics, University of Sofia, BG-1164 Sofia, Bulgaria} \\
$^{9}${Present address: Scuola Superiore Meridionale e INFN, Sezione di Napoli, I-80138 Napoli, Italy} \\
$^{10}${Also at Department of Industrial Engineering, University of Roma Tor Vergata, I-00173 Roma, Italy} \\
$^{11}${Also at Department of Electronic Engineering, University of Roma Tor Vergata, I-00173 Roma, Italy} \\
$^{12}${Also at Universit\`a degli Studi del Piemonte Orientale, I-13100 Vercelli, Italy} \\
$^{13}${Also at Universidad de Guanajuato, 36000 Guanajuato, Mexico} \\
$^{14}${Present address: INFN, Sezione di Perugia, I-06100 Perugia, Italy} \\
$^{15}${Present address: Charles University, 116 36 Prague 1, Czech Republic} \\
$^{16}${Present address: Syracuse University, Syracuse, NY 13244, USA} \\
$^{17}${Present address: Laboratori Nazionali di Frascati, I-00044 Frascati, Italy} \\
$^{18}${Present address: Dipartimento di Fisica dell'Universit\`a e INFN, Sezione di Genova, I-16146 Genova, Italy} \\
$^{19}${Present address: Physics Department, University of Lancaster, Lancaster, LA1 4YB, UK} \\
$^{20}${Also at SLAC National Accelerator Laboratory, Stanford University, Menlo Park, CA 94025, USA} \\
$^{21}${Also at L. N. Gumilyov Eurasian National University, 010000 Nur-Sultan, Kazakhstan} \\
$^{22}${Present address: INFN, Sezione di Pisa, I-56100 Pisa, Italy} \\
$^{23}${Present address: Institute of Nuclear Research and Nuclear Energy of Bulgarian Academy of Science (INRNE-BAS), BG-1784 Sofia, Bulgaria} \\

%% file: NA62_ee_accepted.bbl
\begin{thebibliography}{37}%
\makeatletter
\providecommand \@ifxundefined [1]{%
 \@ifx{#1\undefined}
}%
\providecommand \@ifnum [1]{%
 \ifnum #1\expandafter \@firstoftwo
 \else \expandafter \@secondoftwo
 \fi
}%
\providecommand \@ifx [1]{%
 \ifx #1\expandafter \@firstoftwo
 \else \expandafter \@secondoftwo
 \fi
}%
\providecommand \natexlab [1]{#1}%
\providecommand \enquote  [1]{``#1''}%
\providecommand \bibnamefont  [1]{#1}%
\providecommand \bibfnamefont [1]{#1}%
\providecommand \citenamefont [1]{#1}%
\providecommand \href@noop [0]{\@secondoftwo}%
\providecommand \href [0]{\begingroup \@sanitize@url \@href}%
\providecommand \@href[1]{\@@startlink{#1}\@@href}%
\providecommand \@@href[1]{\endgroup#1\@@endlink}%
\providecommand \@sanitize@url [0]{\catcode `\\12\catcode `\$12\catcode
  `\&12\catcode `\#12\catcode `\^12\catcode `\_12\catcode `\%12\relax}%
\providecommand \@@startlink[1]{}%
\providecommand \@@endlink[0]{}%
\providecommand \url  [0]{\begingroup\@sanitize@url \@url }%
\providecommand \@url [1]{\endgroup\@href {#1}{\urlprefix }}%
\providecommand \urlprefix  [0]{URL }%
\providecommand \Eprint [0]{\href }%
\providecommand \doibase [0]{https://doi.org/}%
\providecommand \selectlanguage [0]{\@gobble}%
\providecommand \bibinfo  [0]{\@secondoftwo}%
\providecommand \bibfield  [0]{\@secondoftwo}%
\providecommand \translation [1]{[#1]}%
\providecommand \BibitemOpen [0]{}%
\providecommand \bibitemStop [0]{}%
\providecommand \bibitemNoStop [0]{.\EOS\space}%
\providecommand \EOS [0]{\spacefactor3000\relax}%
\providecommand \BibitemShut  [1]{\csname bibitem#1\endcsname}%
\let\auto@bib@innerbib\@empty
\bibitem [{\citenamefont {Okun}(1982)}]{Okun}%
  \BibitemOpen
  \bibfield  {author} {\bibinfo {author} {\bibfnamefont {L.~B.}\ \bibnamefont
  {Okun}},\ }\bibfield  {title} {\bibinfo {title} {{Limits of electrodynamics:
  paraphotons?}},\ }\href@noop {} {\bibfield  {journal} {\bibinfo  {journal}
  {Sov. Phys. JETP}\ }\textbf {\bibinfo {volume} {56}},\ \bibinfo {pages} {502}
  (\bibinfo {year} {1982})}\BibitemShut {NoStop}%
\bibitem [{\citenamefont {Holdom}(1986)}]{Holdom}%
  \BibitemOpen
  \bibfield  {author} {\bibinfo {author} {\bibfnamefont {B.}~\bibnamefont
  {Holdom}},\ }\bibfield  {title} {\bibinfo {title} {{Two U(1)'s and $\epsilon$
  charge shifts}},\ }\href
  {https://doi.org/https://doi.org/10.1016/0370-2693(86)91377-8} {\bibfield
  {journal} {\bibinfo  {journal} {Physics Letters B}\ }\textbf {\bibinfo
  {volume} {166}},\ \bibinfo {pages} {196} (\bibinfo {year}
  {1986})}\BibitemShut {NoStop}%
\bibitem [{\citenamefont {Boehm}\ and\ \citenamefont
  {Fayet}(2004)}]{Boehm_2004}%
  \BibitemOpen
  \bibfield  {author} {\bibinfo {author} {\bibfnamefont {C.}~\bibnamefont
  {Boehm}}\ and\ \bibinfo {author} {\bibfnamefont {P.}~\bibnamefont {Fayet}},\
  }\bibfield  {title} {\bibinfo {title} {{Scalar dark matter candidates}},\
  }\href {https://doi.org/https://doi.org/10.1016/j.nuclphysb.2004.01.015}
  {\bibfield  {journal} {\bibinfo  {journal} {Nuclear Physics B}\ }\textbf
  {\bibinfo {volume} {683}},\ \bibinfo {pages} {219} (\bibinfo {year}
  {2004})}\BibitemShut {NoStop}%
\bibitem [{\citenamefont {Pospelov}\ \emph {et~al.}(2008)\citenamefont
  {Pospelov}, \citenamefont {Ritz},\ and\ \citenamefont
  {Voloshin}}]{Pospelov_2008}%
  \BibitemOpen
  \bibfield  {author} {\bibinfo {author} {\bibfnamefont {M.}~\bibnamefont
  {Pospelov}}, \bibinfo {author} {\bibfnamefont {A.}~\bibnamefont {Ritz}},\
  and\ \bibinfo {author} {\bibfnamefont {M.}~\bibnamefont {Voloshin}},\
  }\bibfield  {title} {\bibinfo {title} {{Secluded WIMP dark matter}},\ }\href
  {https://doi.org/https://doi.org/10.1016/j.physletb.2008.02.052} {\bibfield
  {journal} {\bibinfo  {journal} {Physics Letters B}\ }\textbf {\bibinfo
  {volume} {662}},\ \bibinfo {pages} {53} (\bibinfo {year} {2008})}\BibitemShut
  {NoStop}%
\bibitem [{\citenamefont {Feng}\ and\ \citenamefont {Kumar}(2008)}]{Feng_2008}%
  \BibitemOpen
  \bibfield  {author} {\bibinfo {author} {\bibfnamefont {J.~L.}\ \bibnamefont
  {Feng}}\ and\ \bibinfo {author} {\bibfnamefont {J.}~\bibnamefont {Kumar}},\
  }\bibfield  {title} {\bibinfo {title} {{Dark-Matter Particles without
  Weak-Scale Masses or Weak Interactions}},\ }\href
  {https://doi.org/10.1103/PhysRevLett.101.231301} {\bibfield  {journal}
  {\bibinfo  {journal} {Phys. Rev. Lett.}\ }\textbf {\bibinfo {volume} {101}},\
  \bibinfo {pages} {231301} (\bibinfo {year} {2008})}\BibitemShut {NoStop}%
\bibitem [{\citenamefont {Batell}\ \emph {et~al.}(2009)\citenamefont {Batell},
  \citenamefont {Pospelov},\ and\ \citenamefont {Ritz}}]{Batell_PRD2009}%
  \BibitemOpen
  \bibfield  {author} {\bibinfo {author} {\bibfnamefont {B.}~\bibnamefont
  {Batell}}, \bibinfo {author} {\bibfnamefont {M.}~\bibnamefont {Pospelov}},\
  and\ \bibinfo {author} {\bibfnamefont {A.}~\bibnamefont {Ritz}},\ }\bibfield
  {title} {\bibinfo {title} {{Probing a secluded U(1) at $B$ factories}},\
  }\href {https://doi.org/10.1103/PhysRevD.79.115008} {\bibfield  {journal}
  {\bibinfo  {journal} {Phys. Rev. D}\ }\textbf {\bibinfo {volume} {79}},\
  \bibinfo {pages} {115008} (\bibinfo {year} {2009})}\BibitemShut {NoStop}%
\bibitem [{\citenamefont {Reece}\ and\ \citenamefont
  {Wang}(2009)}]{Reece_JHEP_2009}%
  \BibitemOpen
  \bibfield  {author} {\bibinfo {author} {\bibfnamefont {M.}~\bibnamefont
  {Reece}}\ and\ \bibinfo {author} {\bibfnamefont {L.-T.}\ \bibnamefont
  {Wang}},\ }\bibfield  {title} {\bibinfo {title} {{Searching for the light
  dark gauge boson in GeV-scale experiments}},\ }\href
  {https://doi.org/10.1088/1126-6708/2009/07/051} {\bibfield  {journal}
  {\bibinfo  {journal} {{JHEP}}\ }\textbf {\bibinfo {volume} {07}},\ \bibinfo
  {pages} {051} (\bibinfo {year} {2009})}\BibitemShut {NoStop}%
\bibitem [{\citenamefont {Bl\"{u}mlein}\ and\ \citenamefont
  {Brunner}(2014)}]{Blumlein_PLB2014}%
  \BibitemOpen
  \bibfield  {author} {\bibinfo {author} {\bibfnamefont {J.}~\bibnamefont
  {Bl\"{u}mlein}}\ and\ \bibinfo {author} {\bibfnamefont {J.}~\bibnamefont
  {Brunner}},\ }\bibfield  {title} {\bibinfo {title} {{New exclusion limits on
  dark gauge forces from proton Bremsstrahlung in beam-dump data}},\ }\href
  {https://doi.org/10.1016/j.physletb.2014.02.029} {\bibfield  {journal}
  {\bibinfo  {journal} {Physics Letters B}\ }\textbf {\bibinfo {volume}
  {731}},\ \bibinfo {pages} {320} (\bibinfo {year} {2014})}\BibitemShut
  {NoStop}%
\bibitem [{\citenamefont {Blümlein}\ and\ \citenamefont
  {Brunner}(2011)}]{Blumlein_PLB2011}%
  \BibitemOpen
  \bibfield  {author} {\bibinfo {author} {\bibfnamefont {J.}~\bibnamefont
  {Blümlein}}\ and\ \bibinfo {author} {\bibfnamefont {J.}~\bibnamefont
  {Brunner}},\ }\bibfield  {title} {\bibinfo {title} {{New exclusion limits for
  dark gauge forces from beam-dump data}},\ }\href
  {https://doi.org/https://doi.org/10.1016/j.physletb.2011.05.046} {\bibfield
  {journal} {\bibinfo  {journal} {Physics Letters B}\ }\textbf {\bibinfo
  {volume} {701}},\ \bibinfo {pages} {155} (\bibinfo {year}
  {2011})}\BibitemShut {NoStop}%
\bibitem [{\citenamefont {{Cortina Gil}}\ \emph {et~al.}(2023)\citenamefont
  {{Cortina Gil}} \emph {et~al.}}]{Dobrich:2023dkm}%
  \BibitemOpen
  \bibfield  {author} {\bibinfo {author} {\bibfnamefont {E.}~\bibnamefont
  {{Cortina Gil}}} \emph {et~al.} (\bibinfo {collaboration} {NA62
  Collaboration}),\ }\bibfield  {title} {\bibinfo {title} {{Search for dark
  photon decays to $\mu^+\mu^-$ at NA62}},\ }\href
  {https://doi.org/10.1007/JHEP09(2023)035} {\bibfield  {journal} {\bibinfo
  {journal} {{JHEP}}\ }\textbf {\bibinfo {volume} {09}},\ \bibinfo {pages}
  {035} (\bibinfo {year} {2023})}\BibitemShut {NoStop}%
\bibitem [{\citenamefont {Döbrich}\ \emph {et~al.}(2019)\citenamefont
  {Döbrich}, \citenamefont {Ertas}, \citenamefont {Kahlhoefer},\ and\
  \citenamefont {Spadaro}}]{Dobrich_PLB2019}%
  \BibitemOpen
  \bibfield  {author} {\bibinfo {author} {\bibfnamefont {B.}~\bibnamefont
  {Döbrich}}, \bibinfo {author} {\bibfnamefont {F.}~\bibnamefont {Ertas}},
  \bibinfo {author} {\bibfnamefont {F.}~\bibnamefont {Kahlhoefer}},\ and\
  \bibinfo {author} {\bibfnamefont {T.}~\bibnamefont {Spadaro}},\ }\bibfield
  {title} {\bibinfo {title} {{Model-independent bounds on light pseudoscalars
  from rare B-meson decays}},\ }\href
  {https://doi.org/https://doi.org/10.1016/j.physletb.2019.01.064} {\bibfield
  {journal} {\bibinfo  {journal} {Physics Letters B}\ }\textbf {\bibinfo
  {volume} {790}},\ \bibinfo {pages} {537} (\bibinfo {year}
  {2019})}\BibitemShut {NoStop}%
\bibitem [{\citenamefont {{Cortina Gil}}\ \emph {et~al.}(2017)\citenamefont
  {{Cortina Gil}} \emph {et~al.}}]{na62det}%
  \BibitemOpen
  \bibfield  {author} {\bibinfo {author} {\bibfnamefont {E.}~\bibnamefont
  {{Cortina Gil}}} \emph {et~al.} (\bibinfo {collaboration} {NA62
  Collaboration}),\ }\bibfield  {title} {\bibinfo {title} {{The beam and
  detector of the NA62 experiment at CERN}},\ }\href
  {https://doi.org/10.1088/1748-0221/12/05/P05025} {\bibfield  {journal}
  {\bibinfo  {journal} {J. of Instrumentation}\ }\textbf {\bibinfo {volume}
  {12}},\ \bibinfo {pages} {P05025} (\bibinfo {year} {2017})}\BibitemShut
  {NoStop}%
\bibitem [{\citenamefont {Danielsson}\ \emph {et~al.}(2020)\citenamefont
  {Danielsson} \emph {et~al.}}]{Danielsson_2020}%
  \BibitemOpen
  \bibfield  {author} {\bibinfo {author} {\bibfnamefont {H.}~\bibnamefont
  {Danielsson}} \emph {et~al.},\ }\bibfield  {title} {\bibinfo {title} {{New
  veto hodoscope ANTI-0 for the NA62 experiment at CERN}},\ }\href
  {https://doi.org/10.1088/1748-0221/15/07/C07007} {\bibfield  {journal}
  {\bibinfo  {journal} {J. of Instrumentation}\ }\textbf {\bibinfo {volume}
  {15}},\ \bibinfo {pages} {C07007} (\bibinfo {year} {2020})}\BibitemShut
  {NoStop}%
\bibitem [{\citenamefont {Niess}\ \emph {et~al.}(2018)\citenamefont {Niess},
  \citenamefont {Barnoud}, \citenamefont {Cârloganu},\ and\ \citenamefont {{Le
  Ménédeu}}}]{PUMAS}%
  \BibitemOpen
  \bibfield  {author} {\bibinfo {author} {\bibfnamefont {V.}~\bibnamefont
  {Niess}}, \bibinfo {author} {\bibfnamefont {A.}~\bibnamefont {Barnoud}},
  \bibinfo {author} {\bibfnamefont {C.}~\bibnamefont {Cârloganu}},\ and\
  \bibinfo {author} {\bibfnamefont {E.}~\bibnamefont {{Le Ménédeu}}},\
  }\bibfield  {title} {\bibinfo {title} {{Backward Monte-Carlo applied to muon
  transport}},\ }\href
  {https://doi.org/https://doi.org/10.1016/j.cpc.2018.04.001} {\bibfield
  {journal} {\bibinfo  {journal} {Computer Physics Communications}\ }\textbf
  {\bibinfo {volume} {229}},\ \bibinfo {pages} {54} (\bibinfo {year}
  {2018})}\BibitemShut {NoStop}%
\bibitem [{\citenamefont {Allison}\ \emph {et~al.}(2016)\citenamefont {Allison}
  \emph {et~al.}}]{geant4}%
  \BibitemOpen
  \bibfield  {author} {\bibinfo {author} {\bibfnamefont {J.}~\bibnamefont
  {Allison}} \emph {et~al.},\ }\bibfield  {title} {\bibinfo {title} {{Recent
  developments in Geant4}},\ }\href
  {https://doi.org/https://doi.org/10.1016/j.nima.2016.06.125} {\bibfield
  {journal} {\bibinfo  {journal} {Nucl. Instrum. Meth. A}\ }\textbf {\bibinfo
  {volume} {835}},\ \bibinfo {pages} {186} (\bibinfo {year}
  {2016})}\BibitemShut {NoStop}%
\bibitem [{\citenamefont {Junk}(1999)}]{CLsMethod}%
  \BibitemOpen
  \bibfield  {author} {\bibinfo {author} {\bibfnamefont {T.}~\bibnamefont
  {Junk}},\ }\bibfield  {title} {\bibinfo {title} {{Confidence level
  computation for combining searches with small statistics}},\ }\href
  {https://doi.org/https://doi.org/10.1016/S0168-9002(99)00498-2} {\bibfield
  {journal} {\bibinfo  {journal} {Nucl. Instrum. Meth. A}\ }\textbf {\bibinfo
  {volume} {434}},\ \bibinfo {pages} {435} (\bibinfo {year}
  {1999})}\BibitemShut {NoStop}%
\bibitem [{\citenamefont {Cowan}\ \emph {et~al.}(2011)\citenamefont {Cowan},
  \citenamefont {Cranmer}, \citenamefont {Gross},\ and\ \citenamefont
  {Vitells}}]{ProfiledLikelihood}%
  \BibitemOpen
  \bibfield  {author} {\bibinfo {author} {\bibfnamefont {G.}~\bibnamefont
  {Cowan}}, \bibinfo {author} {\bibfnamefont {K.}~\bibnamefont {Cranmer}},
  \bibinfo {author} {\bibfnamefont {E.}~\bibnamefont {Gross}},\ and\ \bibinfo
  {author} {\bibfnamefont {O.}~\bibnamefont {Vitells}},\ }\bibfield  {title}
  {\bibinfo {title} {{Asymptotic formulae for likelihood-based tests of new
  physics}},\ }\href {https://doi.org/10.1140/epjc/s10052-011-1554-0}
  {\bibfield  {journal} {\bibinfo  {journal} {Eur. Phys. J. C}\ }\textbf
  {\bibinfo {volume} {71}},\ \bibinfo {pages} {1554} (\bibinfo {year}
  {2011})}\BibitemShut {NoStop}%
\bibitem [{\citenamefont {Riordan}\ \emph {et~al.}(1987)\citenamefont {Riordan}
  \emph {et~al.}}]{Riordan_E141_PRL1987}%
  \BibitemOpen
  \bibfield  {author} {\bibinfo {author} {\bibfnamefont {E.~M.}\ \bibnamefont
  {Riordan}} \emph {et~al.},\ }\bibfield  {title} {\bibinfo {title} {{Search
  for short-lived axions in an electron-beam-dump experiment}},\ }\href
  {https://doi.org/10.1103/PhysRevLett.59.755} {\bibfield  {journal} {\bibinfo
  {journal} {Phys. Rev. Lett.}\ }\textbf {\bibinfo {volume} {59}},\ \bibinfo
  {pages} {755} (\bibinfo {year} {1987})}\BibitemShut {NoStop}%
\bibitem [{\citenamefont {Davier}\ and\ \citenamefont {{Nguyen
  Ngoc}}(1989)}]{Davier_Orsay_PLB1989}%
  \BibitemOpen
  \bibfield  {author} {\bibinfo {author} {\bibfnamefont {M.}~\bibnamefont
  {Davier}}\ and\ \bibinfo {author} {\bibfnamefont {H.}~\bibnamefont {{Nguyen
  Ngoc}}},\ }\bibfield  {title} {\bibinfo {title} {{An unambiguous search for a
  light Higgs boson}},\ }\href
  {https://doi.org/https://doi.org/10.1016/0370-2693(89)90174-3} {\bibfield
  {journal} {\bibinfo  {journal} {Physics Letters B}\ }\textbf {\bibinfo
  {volume} {229}},\ \bibinfo {pages} {150} (\bibinfo {year}
  {1989})}\BibitemShut {NoStop}%
\bibitem [{\citenamefont {Bl\"{u}mlein}\ \emph {et~al.}(1992)\citenamefont
  {Bl\"{u}mlein} \emph {et~al.}}]{Blumlein_NuCal_IJMA1992}%
  \BibitemOpen
  \bibfield  {author} {\bibinfo {author} {\bibfnamefont {J.}~\bibnamefont
  {Bl\"{u}mlein}} \emph {et~al.},\ }\bibfield  {title} {\bibinfo {title}
  {{Limits on the mass of light (pseudo)scalar particles from Bethe-Heitler
  $e^+e^-$ and $\mu^+\mu^-$ pair production in a proton-iron beam dump
  experiment}},\ }\href {https://doi.org/10.1142/S0217751X9200171X} {\bibfield
  {journal} {\bibinfo  {journal} {International Journal of Modern Physics A}\
  }\textbf {\bibinfo {volume} {07}},\ \bibinfo {pages} {3835} (\bibinfo {year}
  {1992})}\BibitemShut {NoStop}%
\bibitem [{\citenamefont {Andreas}\ \emph {et~al.}(2012)\citenamefont
  {Andreas}, \citenamefont {Niebuhr},\ and\ \citenamefont
  {Ringwald}}]{Andreas_Orsay_PRD2012}%
  \BibitemOpen
  \bibfield  {author} {\bibinfo {author} {\bibfnamefont {S.}~\bibnamefont
  {Andreas}}, \bibinfo {author} {\bibfnamefont {C.}~\bibnamefont {Niebuhr}},\
  and\ \bibinfo {author} {\bibfnamefont {A.}~\bibnamefont {Ringwald}},\
  }\bibfield  {title} {\bibinfo {title} {{New limits on hidden photons from
  past electron beam dumps}},\ }\href
  {https://doi.org/10.1103/PhysRevD.86.095019} {\bibfield  {journal} {\bibinfo
  {journal} {Phys. Rev. D}\ }\textbf {\bibinfo {volume} {86}},\ \bibinfo
  {pages} {095019} (\bibinfo {year} {2012})}\BibitemShut {NoStop}%
\bibitem [{\citenamefont {Lees}\ \emph {et~al.}(2014)\citenamefont {Lees} \emph
  {et~al.}}]{Lees_BaBar_PRL2014}%
  \BibitemOpen
  \bibfield  {author} {\bibinfo {author} {\bibfnamefont {J.~P.}\ \bibnamefont
  {Lees}} \emph {et~al.} (\bibinfo {collaboration} {BaBar Collaboration}),\
  }\bibfield  {title} {\bibinfo {title} {{Search for a Dark Photon in
  ${e}^{+}{e}^{\ensuremath{-}}$ Collisions at BaBar}},\ }\href
  {https://doi.org/10.1103/PhysRevLett.113.201801} {\bibfield  {journal}
  {\bibinfo  {journal} {Phys. Rev. Lett.}\ }\textbf {\bibinfo {volume} {113}},\
  \bibinfo {pages} {201801} (\bibinfo {year} {2014})}\BibitemShut {NoStop}%
\bibitem [{\citenamefont {Batley}\ \emph {et~al.}(2015)\citenamefont {Batley}
  \emph {et~al.}}]{Batley_NA48_PLB2015}%
  \BibitemOpen
  \bibfield  {author} {\bibinfo {author} {\bibfnamefont {J.}~\bibnamefont
  {Batley}} \emph {et~al.},\ }\bibfield  {title} {\bibinfo {title} {{Search for
  the dark photon in $\pi^0$ decays}},\ }\href
  {https://doi.org/https://doi.org/10.1016/j.physletb.2015.04.068} {\bibfield
  {journal} {\bibinfo  {journal} {Physics Letters B}\ }\textbf {\bibinfo
  {volume} {746}},\ \bibinfo {pages} {178} (\bibinfo {year}
  {2015})}\BibitemShut {NoStop}%
\bibitem [{\citenamefont {Banerjee}\ \emph {et~al.}(2020)\citenamefont
  {Banerjee} \emph {et~al.}}]{Banerjee_NA64_2020}%
  \BibitemOpen
  \bibfield  {author} {\bibinfo {author} {\bibfnamefont {D.}~\bibnamefont
  {Banerjee}} \emph {et~al.} (\bibinfo {collaboration} {The NA64
  Collaboration}),\ }\bibfield  {title} {\bibinfo {title} {{Improved limits on
  a hypothetical $X(16.7)$ boson and a dark photon decaying into
  ${e}^{+}{e}^{\ensuremath{-}}$ pairs}},\ }\href
  {https://doi.org/10.1103/PhysRevD.101.071101} {\bibfield  {journal} {\bibinfo
   {journal} {Phys. Rev. D}\ }\textbf {\bibinfo {volume} {101}},\ \bibinfo
  {pages} {071101} (\bibinfo {year} {2020})}\BibitemShut {NoStop}%
\bibitem [{\citenamefont {Abreu}\ \emph {et~al.}(2024)\citenamefont {Abreu}
  \emph {et~al.}}]{Abreu_FASER_PLB2024}%
  \BibitemOpen
  \bibfield  {author} {\bibinfo {author} {\bibfnamefont {H.}~\bibnamefont
  {Abreu}} \emph {et~al.},\ }\bibfield  {title} {\bibinfo {title} {{Search for
  dark photons with the FASER detector at the LHC}},\ }\href
  {https://doi.org/https://doi.org/10.1016/j.physletb.2023.138378} {\bibfield
  {journal} {\bibinfo  {journal} {Physics Letters B}\ }\textbf {\bibinfo
  {volume} {848}},\ \bibinfo {pages} {138378} (\bibinfo {year}
  {2024})}\BibitemShut {NoStop}%
\bibitem [{\citenamefont {Gninenko}(2012)}]{Gninenko_NOMAD_PS191_PRD2012}%
  \BibitemOpen
  \bibfield  {author} {\bibinfo {author} {\bibfnamefont {S.~N.}\ \bibnamefont
  {Gninenko}},\ }\bibfield  {title} {\bibinfo {title} {{Stringent limits on the
  ${\ensuremath{\pi}}^{0}\ensuremath{\rightarrow}\ensuremath{\gamma}X$,
  $X\ensuremath{\rightarrow}{e}^{+}{e}^{\ensuremath{-}}$ decay from neutrino
  experiments and constraints on new light gauge bosons}},\ }\href
  {https://doi.org/10.1103/PhysRevD.85.055027} {\bibfield  {journal} {\bibinfo
  {journal} {Phys. Rev. D}\ }\textbf {\bibinfo {volume} {85}},\ \bibinfo
  {pages} {055027} (\bibinfo {year} {2012})}\BibitemShut {NoStop}%
\bibitem [{\citenamefont {Essig}\ \emph {et~al.}(2011)\citenamefont {Essig},
  \citenamefont {Schuster}, \citenamefont {Toro},\ and\ \citenamefont
  {Wojtsekhowski}}]{Essig_E137_JHEP2011}%
  \BibitemOpen
  \bibfield  {author} {\bibinfo {author} {\bibfnamefont {R.}~\bibnamefont
  {Essig}}, \bibinfo {author} {\bibfnamefont {P.}~\bibnamefont {Schuster}},
  \bibinfo {author} {\bibfnamefont {N.}~\bibnamefont {Toro}},\ and\ \bibinfo
  {author} {\bibfnamefont {B.}~\bibnamefont {Wojtsekhowski}},\ }\bibfield
  {title} {\bibinfo {title} {{An electron fixed target experiment to search for
  a new vector boson $A'$ decaying to $e^+e^-$}},\ }\href
  {https://doi.org/10.1007/JHEP02(2011)009} {\bibfield  {journal} {\bibinfo
  {journal} {{JHEP}}\ }\textbf {\bibinfo {volume} {02}},\ \bibinfo {pages} {9}
  (\bibinfo {year} {2011})}\BibitemShut {NoStop}%
\bibitem [{\citenamefont {Aaij}\ \emph {et~al.}(2018)\citenamefont {Aaij} \emph
  {et~al.}}]{Aaij:2017rft}%
  \BibitemOpen
  \bibfield  {author} {\bibinfo {author} {\bibfnamefont {R.}~\bibnamefont
  {Aaij}} \emph {et~al.} (\bibinfo {collaboration} {LHCb}),\ }\bibfield
  {title} {\bibinfo {title} {{Search for Dark Photons Produced in 13 TeV $pp$
  Collisions}},\ }\href {https://doi.org/10.1103/PhysRevLett.120.061801}
  {\bibfield  {journal} {\bibinfo  {journal} {Phys. Rev. Lett.}\ }\textbf
  {\bibinfo {volume} {120}},\ \bibinfo {pages} {061801} (\bibinfo {year}
  {2018})}\BibitemShut {NoStop}%
\bibitem [{\citenamefont {Konaka}\ \emph {et~al.}(1986)\citenamefont {Konaka}
  \emph {et~al.}}]{Konaka:1986cb}%
  \BibitemOpen
  \bibfield  {author} {\bibinfo {author} {\bibfnamefont {A.}~\bibnamefont
  {Konaka}} \emph {et~al.},\ }\bibfield  {title} {\bibinfo {title} {{Search for
  Neutral Particles in Electron Beam Dump Experiment}},\ }\bibfield
  {booktitle} {\emph {\bibinfo {booktitle} {Proceedings, 23RD International
  Conference on High Energy Physics, JULY 16-23, 1986, Berkeley, CA}},\ }\href
  {https://doi.org/10.1103/PhysRevLett.57.659} {\bibfield  {journal} {\bibinfo
  {journal} {Phys. Rev. Lett.}\ }\textbf {\bibinfo {volume} {57}},\ \bibinfo
  {pages} {659} (\bibinfo {year} {1986})}\BibitemShut {NoStop}%
\bibitem [{\citenamefont {Essig}\ \emph {et~al.}(2010)\citenamefont {Essig},
  \citenamefont {Harnik}, \citenamefont {Kaplan},\ and\ \citenamefont
  {Toro}}]{PhysRevD.82.113008}%
  \BibitemOpen
  \bibfield  {author} {\bibinfo {author} {\bibfnamefont {R.}~\bibnamefont
  {Essig}}, \bibinfo {author} {\bibfnamefont {R.}~\bibnamefont {Harnik}},
  \bibinfo {author} {\bibfnamefont {J.}~\bibnamefont {Kaplan}},\ and\ \bibinfo
  {author} {\bibfnamefont {N.}~\bibnamefont {Toro}},\ }\bibfield  {title}
  {\bibinfo {title} {{Discovering new light states at neutrino experiments}},\
  }\href {https://doi.org/10.1103/PhysRevD.82.113008} {\bibfield  {journal}
  {\bibinfo  {journal} {Phys. Rev. D}\ }\textbf {\bibinfo {volume} {82}},\
  \bibinfo {pages} {113008} (\bibinfo {year} {2010})}\BibitemShut {NoStop}%
\bibitem [{\citenamefont {Ilten}\ \emph {et~al.}(2018)\citenamefont {Ilten},
  \citenamefont {Soreq}, \citenamefont {Williams},\ and\ \citenamefont
  {Xue}}]{Ilten2018}%
  \BibitemOpen
  \bibfield  {author} {\bibinfo {author} {\bibfnamefont {P.}~\bibnamefont
  {Ilten}}, \bibinfo {author} {\bibfnamefont {Y.}~\bibnamefont {Soreq}},
  \bibinfo {author} {\bibfnamefont {M.}~\bibnamefont {Williams}},\ and\
  \bibinfo {author} {\bibfnamefont {W.}~\bibnamefont {Xue}},\ }\bibfield
  {title} {\bibinfo {title} {{Serendipity in dark photon searches}},\ }\href
  {https://doi.org/10.1007/JHEP06(2018)004} {\bibfield  {journal} {\bibinfo
  {journal} {{JHEP}}\ }\textbf {\bibinfo {volume} {06}},\ \bibinfo {pages}
  {004} (\bibinfo {year} {2018})}\BibitemShut {NoStop}%
\bibitem [{\citenamefont {Chang}\ \emph {et~al.}(2017)\citenamefont {Chang},
  \citenamefont {Essig},\ and\ \citenamefont {McDermott}}]{Chang2017}%
  \BibitemOpen
  \bibfield  {author} {\bibinfo {author} {\bibfnamefont {J.~H.}\ \bibnamefont
  {Chang}}, \bibinfo {author} {\bibfnamefont {R.}~\bibnamefont {Essig}},\ and\
  \bibinfo {author} {\bibfnamefont {S.~D.}\ \bibnamefont {McDermott}},\
  }\bibfield  {title} {\bibinfo {title} {{Revisiting Supernova 1987A
  constraints on dark photons}},\ }\href
  {https://doi.org/10.1007/JHEP01(2017)107} {\bibfield  {journal} {\bibinfo
  {journal} {{JHEP}}\ }\textbf {\bibinfo {volume} {01}},\ \bibinfo {pages}
  {107} (\bibinfo {year} {2017})}\BibitemShut {NoStop}%
\bibitem [{\citenamefont {Dorenbosch}\ \emph {et~al.}(1986)\citenamefont
  {Dorenbosch} \emph {et~al.}}]{charm}%
  \BibitemOpen
  \bibfield  {author} {\bibinfo {author} {\bibfnamefont {J.}~\bibnamefont
  {Dorenbosch}} \emph {et~al.} (\bibinfo {collaboration} {CHARM
  Collaboration}),\ }\bibfield  {title} {\bibinfo {title} {{A search for decays
  of heavy neutrinos in the mass range 0.5–2.8 GeV}},\ }\href
  {https://doi.org/https://doi.org/10.1016/0370-2693(86)91601-1} {\bibfield
  {journal} {\bibinfo  {journal} {Physics Letters B}\ }\textbf {\bibinfo
  {volume} {166}},\ \bibinfo {pages} {473} (\bibinfo {year}
  {1986})}\BibitemShut {NoStop}%
\bibitem [{\citenamefont {Sj\"{o}strand}\ \emph {et~al.}(2015)\citenamefont
  {Sj\"{o}strand}, \citenamefont {Ask}, \citenamefont {Christiansen},
  \citenamefont {Corke}, \citenamefont {Desai}, \citenamefont {Ilten},
  \citenamefont {Mrenna}, \citenamefont {Prestel}, \citenamefont {Rasmussen},\
  and\ \citenamefont {Skands}}]{SHORSTRAND_PYTHIA82_CPC2015}%
  \BibitemOpen
  \bibfield  {author} {\bibinfo {author} {\bibfnamefont {T.}~\bibnamefont
  {Sj\"{o}strand}}, \bibinfo {author} {\bibfnamefont {S.}~\bibnamefont {Ask}},
  \bibinfo {author} {\bibfnamefont {J.~R.}\ \bibnamefont {Christiansen}},
  \bibinfo {author} {\bibfnamefont {R.}~\bibnamefont {Corke}}, \bibinfo
  {author} {\bibfnamefont {N.}~\bibnamefont {Desai}}, \bibinfo {author}
  {\bibfnamefont {P.}~\bibnamefont {Ilten}}, \bibinfo {author} {\bibfnamefont
  {S.}~\bibnamefont {Mrenna}}, \bibinfo {author} {\bibfnamefont
  {S.}~\bibnamefont {Prestel}}, \bibinfo {author} {\bibfnamefont {C.~O.}\
  \bibnamefont {Rasmussen}},\ and\ \bibinfo {author} {\bibfnamefont {P.~Z.}\
  \bibnamefont {Skands}},\ }\bibfield  {title} {\bibinfo {title} {{An
  introduction to PYTHIA 8.2}},\ }\href
  {https://doi.org/https://doi.org/10.1016/j.cpc.2015.01.024} {\bibfield
  {journal} {\bibinfo  {journal} {Computer Physics Communications}\ }\textbf
  {\bibinfo {volume} {191}},\ \bibinfo {pages} {159} (\bibinfo {year}
  {2015})}\BibitemShut {NoStop}%
\bibitem [{\citenamefont {Workman}\ \emph {et~al.}(2022)\citenamefont {Workman}
  \emph {et~al.}}]{Workman_PDG_PTEP2022}%
  \BibitemOpen
  \bibfield  {author} {\bibinfo {author} {\bibfnamefont {R.~L.}\ \bibnamefont
  {Workman}} \emph {et~al.} (\bibinfo {collaboration} {Particle Data Group}),\
  }\bibfield  {title} {\bibinfo {title} {{Review of Particle Physics}},\ }\href
  {https://doi.org/10.1093/ptep/ptac097} {\bibfield  {journal} {\bibinfo
  {journal} {Progress of Theoretical and Experimental Physics}\ }\textbf
  {\bibinfo {volume} {2022}},\ \bibinfo {pages} {083C01} (\bibinfo {year}
  {2022})}\BibitemShut {NoStop}%
\bibitem [{\citenamefont {D\"obrich}\ \emph {et~al.}(2019)\citenamefont
  {D\"obrich}, \citenamefont {Jaeckel},\ and\ \citenamefont
  {Spadaro}}]{Dobrich_JHEP2019}%
  \BibitemOpen
  \bibfield  {author} {\bibinfo {author} {\bibfnamefont {B.}~\bibnamefont
  {D\"obrich}}, \bibinfo {author} {\bibfnamefont {J.}~\bibnamefont {Jaeckel}},\
  and\ \bibinfo {author} {\bibfnamefont {T.}~\bibnamefont {Spadaro}},\
  }\bibfield  {title} {\bibinfo {title} {{Light in the beam dump. Axion-Like
  Particle production from decay photons in proton beam-dumps}},\ }\href
  {https://doi.org/10.1007/JHEP05(2019)213} {\bibfield  {journal} {\bibinfo
  {journal} {{JHEP}}\ }\textbf {\bibinfo {volume} {05}},\ \bibinfo {pages}
  {213} (\bibinfo {year} {2019})}\BibitemShut {NoStop}%
\bibitem [{\citenamefont {Verkerke}\ and\ \citenamefont
  {Kirkby}(2003)}]{Verkerke_ROOFIT}%
  \BibitemOpen
  \bibfield  {author} {\bibinfo {author} {\bibfnamefont {W.}~\bibnamefont
  {Verkerke}}\ and\ \bibinfo {author} {\bibfnamefont {D.~P.}\ \bibnamefont
  {Kirkby}},\ }\bibfield  {title} {\bibinfo {title} {{The RooFit toolkit for
  data modeling}},\ }\href@noop {} {\bibfield  {journal} {\bibinfo  {journal}
  {eConf}\ }\textbf {\bibinfo {volume} {C0303241}},\ \bibinfo {pages} {MOLT007}
  (\bibinfo {year} {2003})}\BibitemShut {NoStop}%
\end{thebibliography}%
